# Exploring mechanisms leading to composition errors in monazite (CePO$_4$) analysed with atom probe tomography.

Tom Veret[1], Fabien Delaroche[1], Ivan Blum[1], Jonathan Houard[1], Benjamin Klaes[1], Isabelle Mouton[2], Frederic De-Geuser[2], Anne-Magali Seydoux-Guillaume[3], François Vurpillot[1]

[1]*Groupe de Physique des Materiaux UMR CNRS 6634, CORIA UMR 6614—UFR Sciences Site du Madrillet, Avenue de l'Université — BP 12 76801 Saint Etienne du Rouvray, France*

[2] *Univ. Grenoble Alpes, Grenoble INP, SIMaP, F-38000 Grenoble, France*

[3]*UJM-Saint-Etienne, CNRS, LGL-TPE, F-42023, Saint Etienne, France*

## Abstract :

Monazite (CePO$_4$) is widely used in U-Th-Pb geochronology due to its reliable age determinations, although isotopic disturbances often require nanoscale investigation to better understand the mechanisms at play. Atom Probe Tomography (APT) offers unique capabilities for nanoscale chemical analysis and 3D atomic reconstruction but presents challenges for insulating materials such as CePO$_4$, particularly due to oxygen loss during field evaporation. This study investigates the effects of laser wavelength, energy, metallic coatings and detection device on mass spectrum optimization and compositional accuracy in synthetic CePO$_4$ samples. Results show that shorter laser wavelengths (260 nm) enhance peak resolution, particularly when combined with advanced reflectron configurations, as demonstrated with the LEAP 6000 XR. Chromium coatings further improve thermal dissipation and reduce noise levels. However, compositional measurements reveal systematic underestimation of oxygen and overestimation of P and Ce, likely influenced by preferential low-field element evaporation. These findings highlight the need to carefully tune experimental parameters to mitigate quantification biases and enhance the reliability of APT analyses for geological materials.

## 1) Introduction

Monazite is an accessory orthophosphate mineral (light rare earth element phosphate: LREEPO$_4$), one of the mostly used in U-Th-Pb geochronology across various geological contexts. It generally yields concordant dates that can be directly interpreted as the age of the (last) crystallization event (Parrish 1990; Harrison, Catlos et Montel 2002). However, discordant dates are frequent and indicate disturbances in the isotopic systems after crystallization (Seydoux-Guillaume et al. 2018).

Complex compositional zoning is often observed in monazite grains, attesting to successive crystallization events. These zones can be dated individually using in situ microanalytical methods such as electron probe microanalysis (EPMA), secondary ion mass spectrometry (SIMS and nanoSIMS), or laser ablation inductively coupled plasma mass spectrometry (LA-ICP-MS). Such techniques enable the reconstruction of the reaction history of the host rock ( Fougerouse et al. 2020; Seydoux-Guillaume et al. 2018)

Recent studies, however, highlight the necessity for nanoscale characterization to constrain the mechanisms leading to disturbances in the U-Th-Pb systems (Seydoux-Guillaume et al. 2019; Turuani et al 2023; 2024; Fougerouse et al. 2021; Verberne et al. 2024). As spatial resolution of conventional techniques is limited with generally spot diameters ranging from 5 to 35 µm and a minimum analytical volume of approximately 9 µm³ for nanoSIMS (Yang et al. 2012), nanogeochronology using atom probe tomography (APT) has already proved to be a solution to significantly improve isotopic dating and therefore geological interpretation (Turuani et al. 2022; 2024).

APT is an advanced tool for atomic-scale material analysis, capable of elemental analysis and 3D atom-by-atom reconstruction of samples (Müller et al. 1968). This technique was initially developed for metal analysis and is based on time-of-flight mass spectrometry, achieved by field-triggered atom-by-atom evaporation of a sample of interest. In the 2000s, the development of ultra-short pulsed lasers significantly expanded its application to semiconductors and insulating materials. This breakthrough is known as laser-assisted atom probe tomography (La-APT) (Blavette et al. 2008).

The interest in this technique for geological and cosmochemical applications is growing, supported by promising results obtained on various minerals (Reddy et al. 2020). However, recently some authors revealed the complexity of doing quantification with this technique and point to significant differences in composition and isotopic ratio measurements compared to conventional methods, or when analyzing samples of known composition (Seydoux-Guillaume et al. 2019; D. Fougerouse et al. 2018; Fougerouse et al. 2020; Gopon et al. 2022; Exertier et al. 2018). In the literature, these discrepancies are most often associated with oxygen losses during analysis, underlining the challenge of obtaining reliable elemental and isotopic quantification for insulating geological materials (Santhanagopalan et al. 2015; Cappelli et al. 2021).

Even though composition biases are increasingly being considered. Certain of the previously cited studies pay little attention to compositional biases caused by preferential losses, primarily because the elements and isotopes used for dating methods do not appear to be directly affected. However, as demonstrated in Takahashi's 2022 work (Takahashi et al. 2022) on nitrides, the preferential loss of an element during APT analysis could signal a broader loss mechanism, potentially impacting the measurement of the composition or isotopic ratios of other elements. It is therefore crucial to adopt a rigorous methodology to avoid potential systematic measurement biases or, at the very least, to try to understand the physical mechanisms driving these losses.

This study investigates how instrument configuration, experimental parameters, and sample preparation affect mass spectrum resolution and the resulting composition. First, we will investigate the impact of laser wavelength and energy and the use of nanometer-scale metallic coatings (Schwarz et al. 2024) on the specimen surface for straight flight path probes. Next, we analyze these parameters on instruments with a curved ion flight path (reflectron), designed to enhance mass resolving power and reduce background noise.

## 2) Sample and methods

### Samples

The synthesis of $CePO_4$ single crystals is achieved using sol-gel and flux growth methods, as described in (Gardés et al. 2006). These methods produce highly pure single crystals with very high homogeneity on scales larger than those of atom probe specimens, enabling the analysis of individual grain fraction. Various reactants are utilized in these processes, including $Ce_2O_3$, $HNO_3$, $H_3PO_4$, and $NH_4OH$ for the sol-gel method, as well as $Li_2WO_4$-$WO_3$ or $Li_2MoO_4$-$MoO_3$ for flux growth. While these methods are effective, they may introduce slight contaminations into the final crystals. Theses crystal has the same structure as Monazite, which has a monoclinic crystal structure (space group $P2_1/n$). It consists of alternating REE-O polyhedra (REE = rare-earth elements like Ce, La, Nd) and $PO_4$ tetrahedra (Clavier et al. 2011).

## APT specimen preparation

The atom probe samples were prepared following a conventional focused ion beam (FIB) protocol (Prosa and Larson 2017), using a dual-beam scanning electron microscope (Thermofisher Galium 5UX). The protocol description is provided in (Lefebvre-Ulrikson 2016). A 10 nm carbon coating was applied to the sample before FIB milling to limit charging effects. The deposit was carried out using a Precision Etching and Coating System (PECS) from Gatan at a beam energy of 10 keV. The samples were mounted on pre-sharpened tungsten needles, prepared using the electro-polishing method (Lefebvre-Ulrikson 2016) using a 3% NaOH solution, to allow for subsequent shaping and manipulations. The final specimen-shaped specimens had a curvature diameter ranging from 40 to 80 nm.

Samples labelled with "Cr5" mention are prepared by adding a thin Cr coating of a few nanometers to the surface of the specimen. The purpose of this coating is to achieve a higher sample yield in wide bandgap materials, likely due to the mechanical stabilization provided by the layer. This process results in an improved field of view and enhanced mass spectrum resolution, with reduced background noise, possibly due to better thermal conductivity and/or laser energy absorption by the metallic coating (Larson et al. 2013; Schwarz et al., s. d.; Arnoldi et al. 2015). The metal coatings are applied using a PECS. Chromium (Cr) coatings are deposited via ion sputtering. The sample, in tip form, is positioned under the Cr target at a 45° angle with a 10 RPM rotation. The coating is carried out at 5 keV to achieve a 5 nm thickness. In practice, the chromium deposition results in the formation of a CrO layer on the sample surface rather than pure metallic chromium. The term "chromium coating" is thus used here as a simplification to designate these samples.

APT/FIM experiment and Data Processing

The experiments were conducted using several laser-based atom probes to compare the advantages of each instrument's specific features. The first was a Photonic Atom Probe (PAP) with a 260 nm laser wavelength and a nominal flight path of 100 mm (Figure 1.A) (Houard et al. 2020). This instrument is also equipped with a next-generation detector featuring fast acquisition electronics and a multi-hit detection algorithm, which reduces detection losses (Costa et al. 2012; Ndiaye et al. 2023). The second was a Local Electrode Atom Probe (LEAP) 5000 XS with a 355 nm laser wavelength and a nominal flight path of 100 mm (Figure 1.B). Both instruments are housed at the GPM (Groupe de Physique des Matériaux) of the University of Rouen, France. The third instrument is a Local Electrode Atom Probe LEAP 6000 XR with a 260 nm laser wavelength and a curved ion flight path (reflectron) was used (Figure 1.C), housed at the SIMAP at the University of Grenoble-Alpes, France. This configuration improves mass resolving power and reduces background noise. Note that all experiments performed with this instrument were conducted in laser mode only. Schematics of the various instruments are shown in Figure 1 and the different configurations, along with the main parameters used for the experiments discussed in this work, are summarized in Table 1.

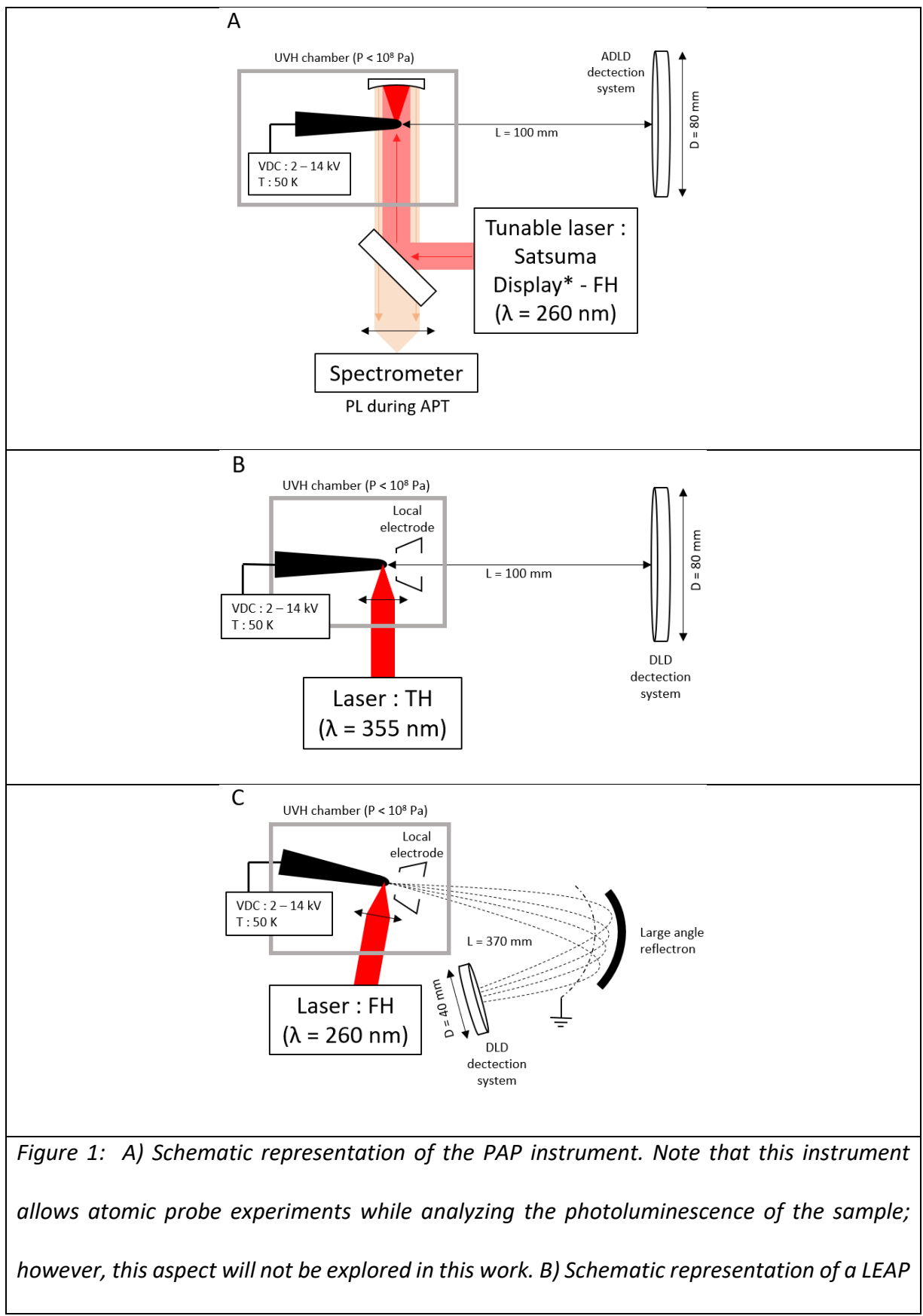

*Figure 1: A) Schematic representation of the PAP instrument. Note that this instrument allows atomic probe experiments while analyzing the photoluminescence of the sample; however, this aspect will not be explored in this work. B) Schematic representation of a LEAP*

*5000 XS from the company CAMECA ©. C) Schematic representation of a LEAP 6000 XR from the company CAMECA ©.*

*The main features of these instruments and the experimental parameters common to the experiments discussed in this work are summarized in Table 1.*

*Table 1: Summary of the various characteristics of the atom probes used in this work, highlighting the key differences between each instrument.*

| Atom Probe | PAP | LEAP 5000 XS | LEAP 6000 XR |
|---|---|---|---|
| Wavelength (nm) | 260 | 355 | 260 |
| Flight path | Strait | Strait | Reflectron |
| Fligth lenth (mm) | 100 | 100 | 370 |
| Spot diameter (µm) | 1 | 2 | 2 |
| Local electrode | No | Yes | Yes |
| Detection system | ADLD | DLD | DLD |
| Detector diameter (mm) | 80 | 80 | 40 |
| Temperature (K) | 50 | 50 | 50 |

Data processing for the LEAP instruments was initially performed using AP Suite software (version 6.3), developed by Cameca® for mass spectrum reconstruction and optimization. The PAP data (reconstruction, spectrum optimization) and all mass spectrum processing for composition measurement were performed using MoDat software (*IDDN.FR.001.140011.000.S.P.2025.000.31235*), developed by the GPM. The "ranging" of the mass spectra for composition measurement was done manually and tailored to each spectrum. All mass spectra were manually processed to account for changes in peak shape and ensure consistent treatment across analyses. Peak ranges were defined to include the full thermal tail up to the background level or the next peak. Although slight adjustments were sometimes required due to variations in evaporation conditions or the appearance of additional peaks, we consider that these effects were minor compared with the compositional biases caused by evaporation mechanisms. As all datasets were processed by the same operator, inter-user variability was eliminated. Previous studies (Hudson et al. 2011; Haley et al. 2015) have shown that automated ranging methods mainly reduce user variance without improving average accuracy. Therefore, the impact of ranging on composition is considered secondary in this work. However, for more complex materials such as natural monazites containing numerous trace elements, automated ranging approaches could become significantly more relevant. For each experiment, the direct voltage (VDC) was allowed to vary freely to maintain a constant detection rate (DR). All measured compositions were compared to the stoichiometric composition of the samples: $CePO_4$ (16.66 At%;16.67 At%;66,67 At%).

An experiment was conducted using a field ion microscope (FIM) to study the behavior of surface atoms during evaporation on a $CePO_4$ sample with a Cr coating. The FIM is the predecessor of the atom probe tomography and shares the same fundamental components. The main differences lie in the detector, which is replaced by a CCD camera, and the use of an imaging gas. The imaging gas is introduced into the analysis chamber at low pressure (~$10^{-5}$ mbar). The sample is subjected to a continuous voltage, generating the necessary electric field to trigger field ionisation of the imaging gas atoms. The emitted ions are then collected by the detection system, forming an image that reveals the atomic-scale structure of the sample surface at a given moment in time. The FIM operates based on gas field ionization (using helium, neon, or hydrogen in this case) and field evaporation of the sample (Müller 1951).

The results were compared to the $CePO_4$ structure (Boatner 2002; Ni et al. 1995), which is schematically illustrated in Figure 7.A. The experimental setup consists of a FIM housed at the GPM laboratory of the University of Rouen, as described in various studies, including the work of B. Klaes (Klaes et al. 2021).

Laser Assisted-Atom Probe Tomography (LA-APT) Theory

The interaction between the specimen and the laser is therefore critical in the ion evaporation mechanism. Here, we consider that the primary process of laser-assisted evaporation is the absorption of laser energy, which generates a thermal pulse and a rapid increase in the sample temperature (T) (Vurpillot et al. 2009; Vella 2013; Kelly 2014). When a semiconductor or dielectric sample is analysed in LA-APT, its absorption properties depend on its band structure (affected by the electric field, doping from FIB preparation, and defects) as well as the shape of the sample, particularly the parts where dimensions are smaller than the laser wavelength (Bogdanowicz 2018). Indeed, in the presence of a strong electric field, the band gap at the surface of a semiconductor sample is reduced, altering its optical properties. Damage caused by FIB preparation also creates an amorphous layer (compared to the crystalline structure of the specimen core), which enhances absorption efficiency and, consequently, heating of the specimen (Silaeva et al. 2014; Bogdanowicz 2018).

The energy of incident photons then exceeds the band gap value. In this case, the penetration depth of the light becomes very small, and all the photon energy is absorbed at the surface. Heating only the surface of the specimen favours the evaporation of surface atoms, particularly at positions where the electric field is strongest, i.e., at the apex of the specimen and the border of atomic terraces (Lefebvre-Ulrikson 2016).

For semiconductor or dielectric materials, laser pulse absorption consists of two contributions: surface absorption and volume absorption. We may note that pulse shaping by light focusing is required to reduce bulk absorption size and improve mass spectrum (Bunton et al. 2007). Surface absorption leads to rapid (thermal) evaporation and well-resolved mass peaks in the time-of-flight spectrum. Volume absorption may cause delayed evaporation, resulting in thermal tails and potentially limiting mass resolution in APT. To achieve optimal evaporation conditions, surface absorption is favoured to improve mass spectrum resolution by sharpening peaks and reducing the length of thermal tails. The mass spectra of monazites are complex and present a forest of peaks from the evaporation of small molecules. Maximizing peak resolution is therefore crucial for accurate composition measurement (Fougerouse et al. 2020) Additionally, confining heating to the surface of the specimen allows the volume to serve as a "thermal sink," reducing the sample's cooling time and, consequently, the thermal tail from residual evaporation. This phenomenon is also influenced by the shape of the specimen. Indeed, the larger the cone angle, the faster the cooling will occur. (Arnoldi et al. 2014; Kumar et al. 2024).

## 3) Result

To measure the composition of a sample in APT, it is essential to focus on the mass-to-charge spectrum. The first step is identifying the peaks in this spectrum. In the case of monazites, the mass spectrum shows a significant number of peaks resulting from the evaporation of molecular ions. During the identification process, it is crucial to detect peaks that may overlap, meaning two elemental species share the same mass-to-charge ratio. Overlap is one of the primary causes of data loss in APT. In all the analyses presented below, peak identification and ranging were manually performed. One of the primary objectives of this work is to optimize the mass spectrum based on the analysis parameters. The goal is to obtain spectra that are as easy to process as possible while minimizing losses, whether caused by evaporation or overlap. Figure 2 shows the mass spectrum of a $CePO_4$ sample obtained using LA-APT, highlighting its main constituent peaks.

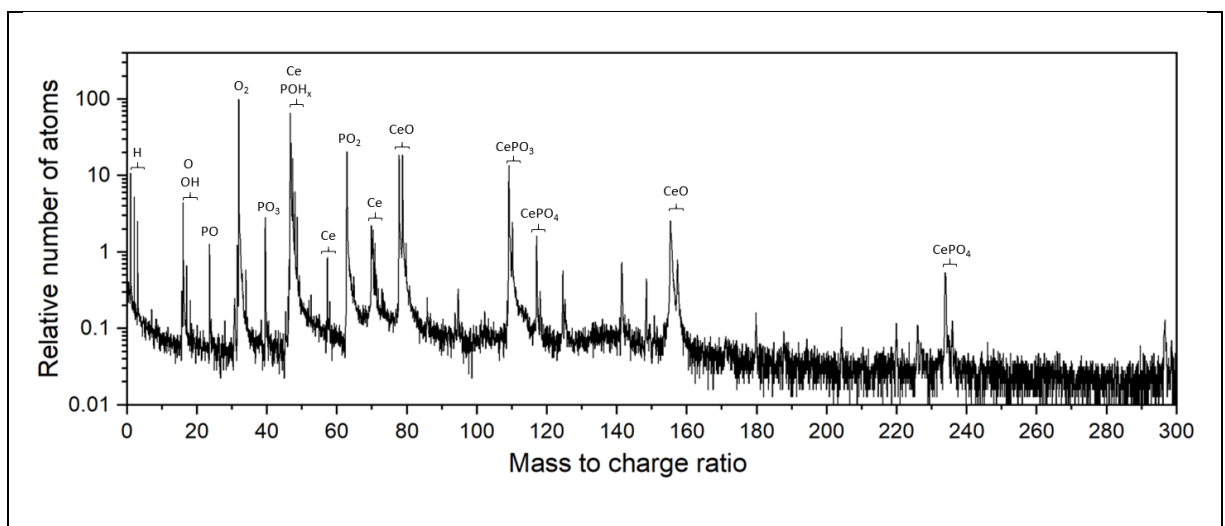

*Figure 2: Mass spectrum of a $CePO_4$ sample obtained using the LEAP 5000 XS at LPE = 2 pJ, T = 50 K, and F = 200 kHz with the main peaks labeled.*

Laser wavelength influence

Two uncoated needles of $CePO_4$ where analysed using the LEAP 5000 XS (355 nm) and PAP (260 nm) under similar conditions: laser pulse energy (LPE) of 100 pJ, detection rate (DR) of 0.1%, and base temperature (T) of 50 K. It is evident that comparing data from two different instruments is complex and involves numerous uncertainties. It is important to recall that both the PAP and the LEAP 5000 XS are straight-flight atom probes with similar flight path lengths and detector sizes (DLD anodes), resulting in comparable fields of view. However, the signal processing technologies used at the detector output differ: the LEAP 5000 XS employs DLD technology, while the PAP uses ADLD, which impacts the detection of multi-hit events. Additionally, beyond the wavelength differences, the laser focus unit also varies, altering the spot size. Nevertheless, it is possible to qualitatively compare these datasets to understand the differences induced by the specific features of each instrument on the mass spectra. Figure 3.A & B presents a segment of the mass spectrum of $CePO_4$ acquired with the LEAP 5000 XS (in red) and PAP (in black). The mass spectrum from the PAP exhibits a noise level approximately ten times higher than that of the LEAP 5000 XS (the origin of which remains uncertain), which hinders the detection of some minor peaks, particularly within the thermal tails of the main peaks. The main peak resolution ($O_2$ 32 Da) at 50%, 10%, and 1% is 355, 168, and 43 for the LEAP 5000 XS, respectively, and 400, 128, and 16 for the PAP instrument.

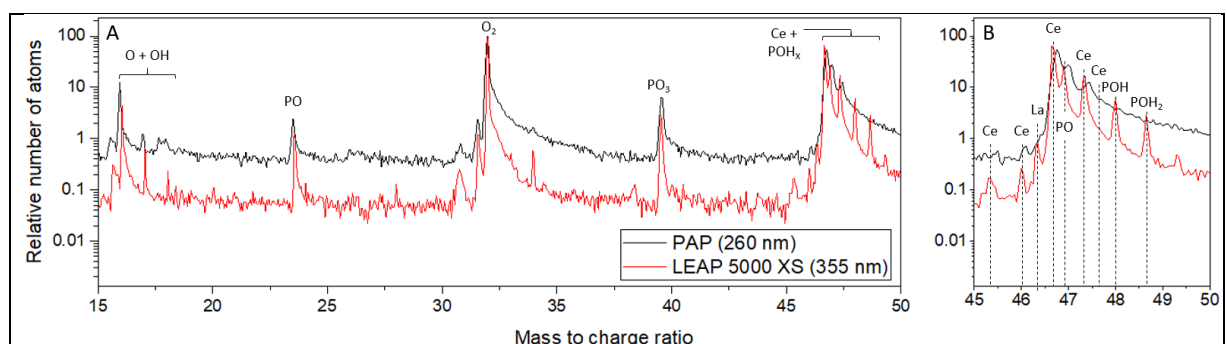

*Figure 3: A) $CePO_4$ mass to charge spectrum obtained using the PAP (Black) and a LEAP 5000 XS (Red) at a 2 pJ laser energy. B) Focus on the area between 45 and 50 Da.*

## Laser pulse energy (LPE) influence

Two CePO$_4$ needles were analyzed using the LEAP 5000 XS and the PAP instrument at different laser energies. Figure.4 A & B presents the mass-to-charge spectra obtained at 100 pJ and 2 pJ on each instrument. For the LEAP 5000 XS analysis, the resolution of the main O$_2$ peak (32 Da) at 50%, 10%, and 1% is 319, 152, and 43, respectively, at 100 pJ, compared to 355, 168, and 49 at 2 pJ. For the PAP instrument analysis, the resolution of the main O$_2$ peak (32 Da) at 50%, 10%, and 1% is 188, 60, and 15, respectively, at 100 pJ, compared to 400, 128, and 16 at 2 pJ. We can also observe a higher background noise level at lower laser energy (or higher electric field) for both instruments, while the thermal tails are less pronounced under these conditions, contributing to an improvement in resolution.

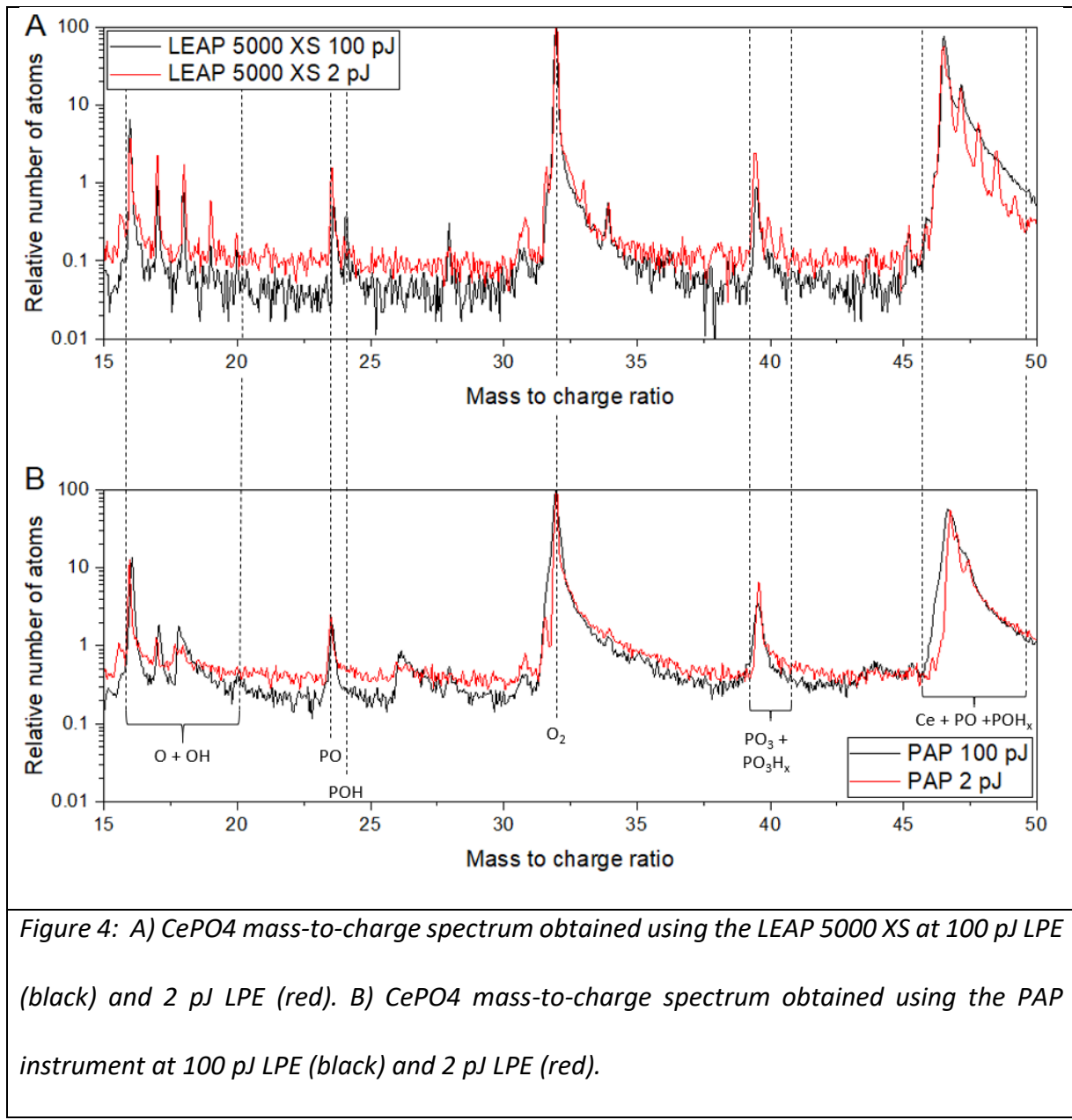

*Figure 4: A) CePO4 mass-to-charge spectrum obtained using the LEAP 5000 XS at 100 pJ LPE (black) and 2 pJ LPE (red). B) CePO4 mass-to-charge spectrum obtained using the PAP instrument at 100 pJ LPE (black) and 2 pJ LPE (red).*

Chromium coating

One method used to reduce background noise is metallic coating through ion sputtering. A 5 nm layer of Chromium (Cr) is added to the specimen after preparation. The presence of a conductive layer on the sample's surface helps reduce bulk heating while enhancing thermal energy dissipation, as the Cr layer is a better thermal electric conductor than Monazite.

Moreover, Cr is also a better electrical conductor than monazite, which helps minimize charge loss between the specimen base, where the potential is applied, and the specimen's apex. This also enhances the longevity of the specimens while improving absorption for certain wavelengths.

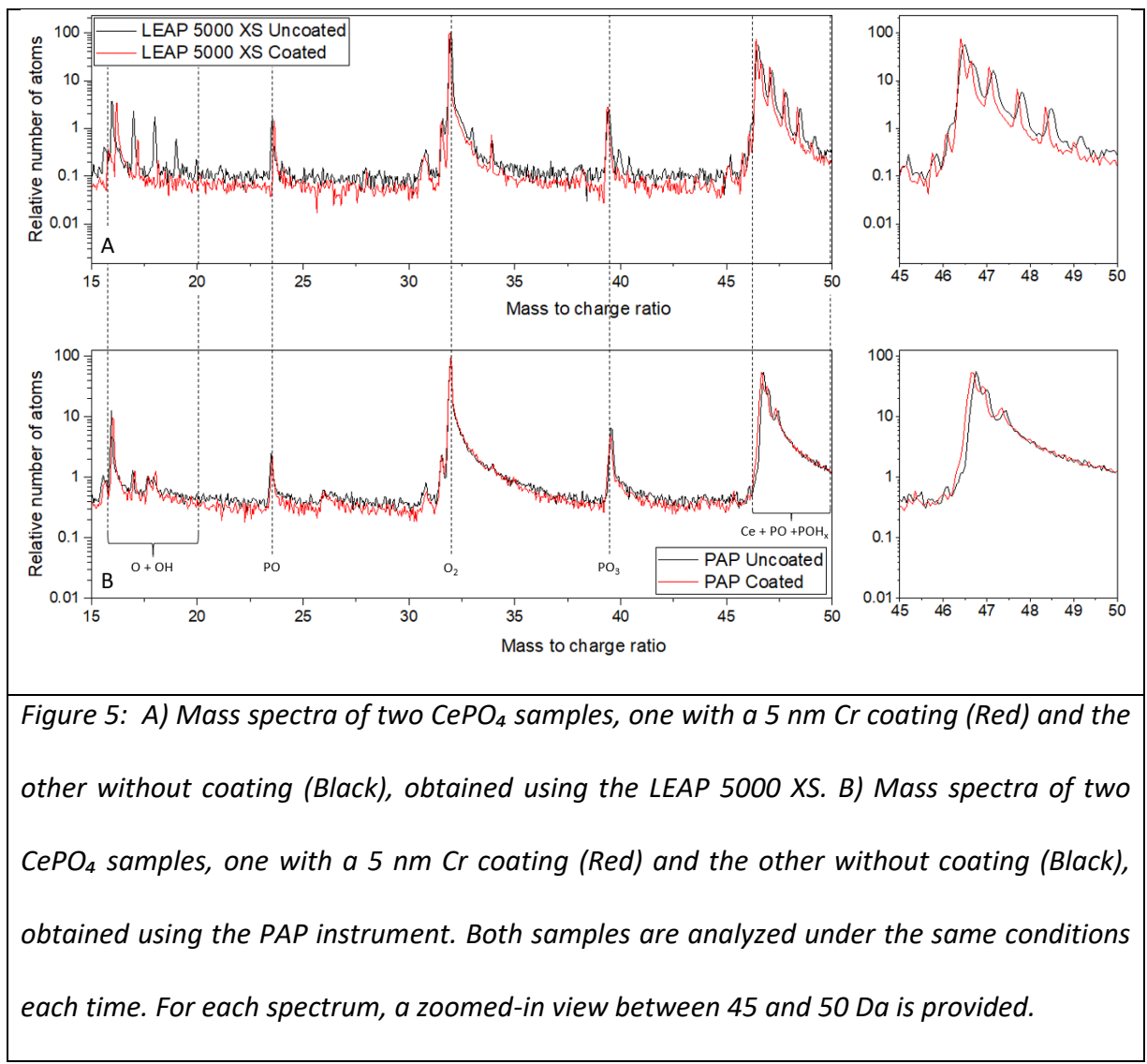

*Figure 5: A) Mass spectra of two CePO$_4$ samples, one with a 5 nm Cr coating (Red) and the other without coating (Black), obtained using the LEAP 5000 XS. B) Mass spectra of two CePO$_4$ samples, one with a 5 nm Cr coating (Red) and the other without coating (Black), obtained using the PAP instrument. Both samples are analyzed under the same conditions each time. For each spectrum, a zoomed-in view between 45 and 50 Da is provided.*

Figure 5.A presents the mass spectrum of two CePO$_4$ samples: one with a 5 nm Cr coating (red) and the other without any treatment obtained with the LEAP 5000 XS at 2pJ LPE. A significant reduction in background noise is observed across the entire mass spectrum. Additionally, a change in peak shape is noticeable, especially in the main peaks, with the coated sample

showing a reduced thermal tail, indicating faster cooling. This leads to an overall improvement in peak resolution throughout the mass spectrum. For the $O_2$ peak (32 Da), the resolution at 50, 10, and 1% is 355, 168, and 43 for the uncoated sample, compared to 639, 319, and 91 for the sample with a 5 nm Cr coating. Figure 5.B shows the same comparison but obtained with the PAP instrument again at 2 pJ LPE. The reduction in background noise is significantly less pronounced, and no changes in the shape of the mass spectra can be observed. The resolution of the main $O_2$ peak (32 Da) at 50%, 10%, and 1% remains unchanged between the two samples, with values of 400, 110 and 18 for the coated specimen and 400, 128, 16 for the uncoated specimen, respectively. The differences in peak heights observed between 16 and 19 Da in Figure 5.A indicate that the Cr coating could, in certain situations, reduce the generation of hydrides during evaporation, thereby minimizing the number of parasitic peaks.

Reflectron

To enhance the resolution of mass spectra, modern APT instruments are equipped with an energy compensation system called a reflectron. This device uses a static electric field in a specific region of the ion path to reverse their direction. The flight path is thus extended and becomes dependent on the energy deficit, enabling both temporal and spatial focusing of the ions. Figure 6.A & B illustrates the difference between spectra obtained with the PAP (straight flight path) and a LEAP 6000 XR (reflectron) on two $CePO_4$ samples. The spectrum from the LEAP 6000 XR shows significantly higher peak resolution and lower noise levels. Focusing on the resolution of the main $O_2$ peak (32 Da) at 50, 10, and 1%, the values are 400, 128, and 16 for the PAP, compared to 1603, 641, and 291 for the LEAP 6000 XR. Both experiments were conducted under the same base temperature and laser pulse energy (LPE) conditions. The spectrum obtained with the LEAP 5000 XS is included for comparison even if the wavelength is different. The differences in the shape of the $O_2$ peaks shown in Figure 6.B indicate that the reflectron, by increasing the flight path and flight time, improves the instrument's mass-resolving power.

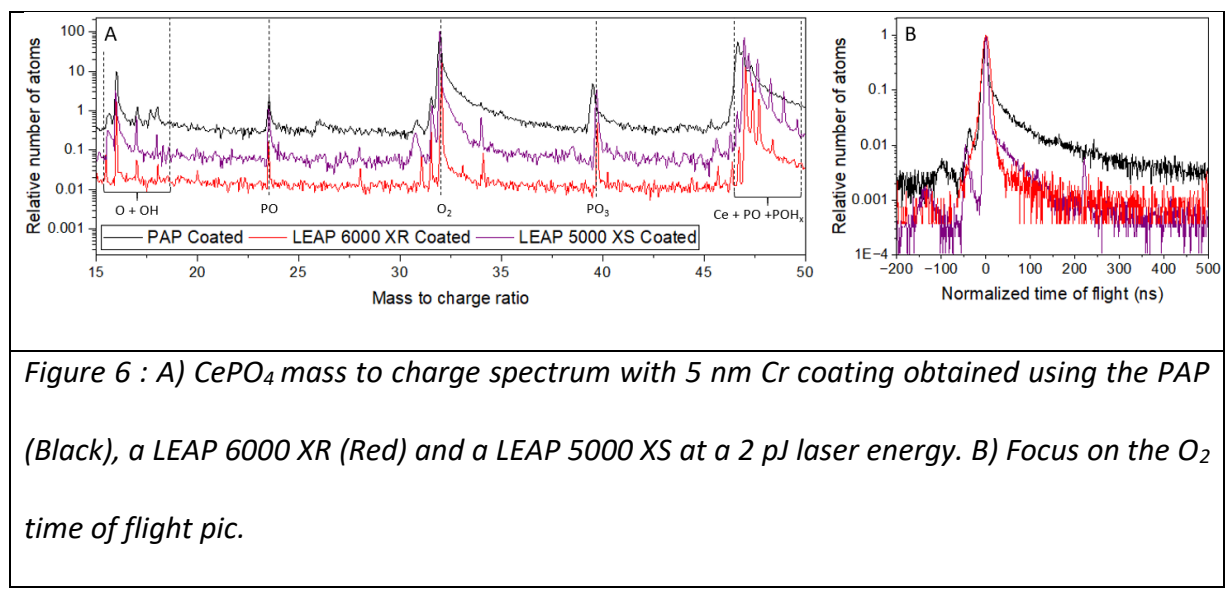

*Figure 6 : A) CePO$_4$ mass to charge spectrum with 5 nm Cr coating obtained using the PAP (Black), a LEAP 6000 XR (Red) and a LEAP 5000 XS at a 2 pJ laser energy. B) Focus on the $O_2$ time of flight pic.*

Field ion microscopy

The FIM experiment was conducted at 60 K to minimize thermal agitation, using a gas mixture of He, Ne, in equal proportion and a small amount of $H_2$ (~$5 \times 10^{-6}$ mbar). The obtained images, an example of which is shown in Figure 7.B, reveal surface atoms (each imaged as a white dot) that appear disorganized yet homogeneous compared to the theoretical structure represented in Figure 7.A The brighter outer ring corresponds to CrO from the previously mentioned coating. All acquired images were similar to the presented picture. No crystallographic features, such as pole structures or variations in contrast over the tip apex, were observed during field evaporation. The CrO layer was continuously observed during the process showing the homogeneous deposition on the complete surface of the monazite. Atoms just before evaporation were imaged as bright spots, which indicated a local field close or a bit lower from the best image field of neon (~35 V/nm). The CrO layer is imaged slightly brighter, indicating an evaporation field of the Monazite slighter lower. We may note than chromia evaporation field under high laser energy pulsing was estimated exceeding 23 V/nm (Jakob et al. 2024). Considering that evaporation is here in DC mode at 60K, it gives us an evaporation field for Monazite in the range 20-30 V/nm.

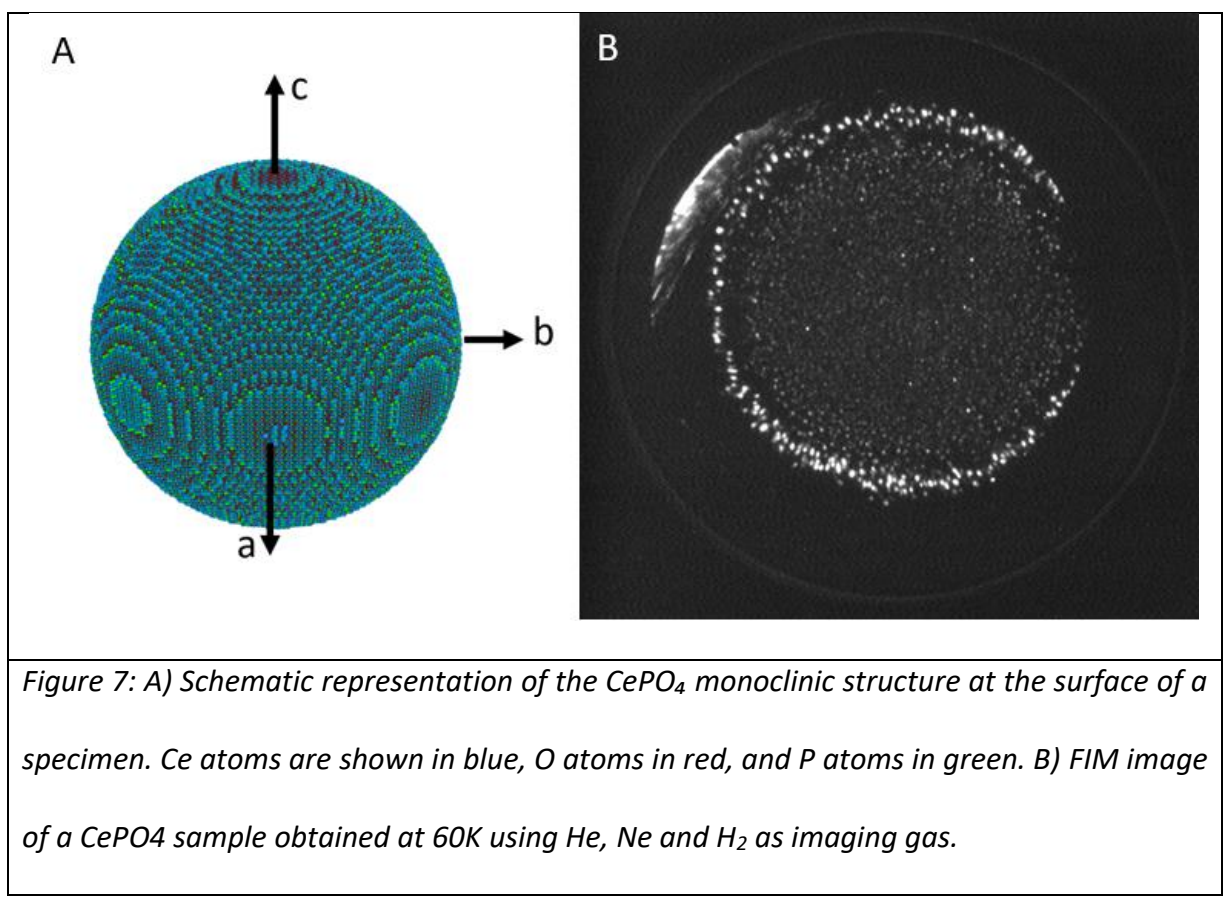

Figure 7: A) Schematic representation of the CePO$_4$ monoclinic structure at the surface of a specimen. Ce atoms are shown in blue, O atoms in red, and P atoms in green. B) FIM image of a CePO4 sample obtained at 60K using He, Ne and H$_2$ as imaging gas.

## Stoichiometry of CePO$_4$ in APT

The composition measurement of experiments done for this work are presented in Table 2. The concentration of elements measured by APT is compared with the theoretical stoichiometry: Ce (16.66%); P (16.67%); O (66,67%). For all experiments, Ga traces resulting from FIB preparation are not considered for calculation. Similarly, if the sample is coated with Cr, it is excluded from the composition calculation, such as hydrogen, which mainly originates from the analysis chamber.

Experiments consistently reveal an oxygen deficit, leading to an overestimation of other elements across all instruments and analysis conditions. However, a nearly systematic increase in measured oxygen content is observed when the LPE (laser pulse energy) is reduced (electric field increased), particularly in the LEAP 6000 XR experiments. This instrument shows the most significant variation, also causing notable changes in the phosphorus (P) and cerium (Ce) composition.

In all experiments, a non-negligible amount of hydrogen (H) is detected, likely originating from the chamber. At low LPE, the measured hydrogen content appears directly correlated with an increase in noise levels.

For experiments conducted on needles without metallic deposition using the LEAP 5000 XS and the PAP, higher noise levels are observed at high LPE settings (100 and 80 pJ). Noise decreases at intermediate LPE levels, only to rise again at low LPE (5 and 2 pJ) probably caused by DC field evaporation at high electric field.

*Table 2: Chemical analysis of atom probe samples prepared from synthetic monazite samples described in Table 1. Noise is measured by accounting for the atomic percent (at%) outside of the defined ranges. It is important to note that trace elements and hydrogen are not included in the composition calculation. The hydrogen level is still expected.*

| APT | Coating | LPE (pJ) | Hydrogen level (at%) H | ±2σ | Concentration relacculated with H excluded (at%) Total | P | ±2σ | O | ±2σ | Ce | ±2σ | Noise |
|---|---|---|---|---|---|---|---|---|---|---|---|---|
| PAP (260 nm) | None | 100.00 | 0.62 | 0.02 | 99.38 | 15.15 | 0.11 | 65.68 | 0.15 | 19.17 | 0.12 | 43.19 |
| | | 80.00 | 1.29 | 0.04 | 98.71 | 12.75 | 0.12 | 64.90 | 0.18 | 22.35 | 0.16 | 46.00 |
| | | 50.00 | 1.01 | 0.03 | 98.99 | 18.63 | 0.10 | 64.46 | 0.13 | 16.91 | 0.10 | 31.45 |
| | | 20.00 | 1.07 | 0.03 | 98.93 | 18.39 | 0.12 | 64.76 | 0.14 | 16.85 | 0.11 | 33.43 |
| | | 10.00 | 1.22 | 0.03 | 98.78 | 18.48 | 0.11 | 65.61 | 0.13 | 15.92 | 0.10 | 40.71 |
| | | 5.00 | 1.40 | 0.04 | 98.60 | 18.39 | 0.14 | 65.64 | 0.18 | 15.97 | 0.14 | 47.39 |
| | | 2.00 | 1.33 | 0.04 | 98.67 | 18.44 | 0.13 | 66.00 | 0.16 | 15.57 | 0.12 | 45.84 |
| PAP (260 nm) | Cr 5nm | 100.00 | 1.59 | 0.03 | 98.41 | 20.08 | 0.10 | 62.05 | 0.12 | 17.87 | 0.10 | 27.54 |
| | | 80.00 | 1.41 | 0.02 | 98.59 | 20.14 | 0.08 | 62.18 | 0.10 | 17.68 | 0.08 | 27.29 |
| | | 50.00 | 1.35 | 0.03 | 98.65 | 20.09 | 0.10 | 62.99 | 0.12 | 16.92 | 0.09 | 28.49 |
| | | 20.00 | 1.34 | 0.02 | 98.66 | 20.00 | 0.08 | 63.71 | 0.10 | 16.29 | 0.08 | 29.68 |
| | | 10.00 | 1.39 | 0.03 | 98.61 | 19.33 | 0.10 | 64.72 | 0.12 | 15.95 | 0.09 | 32.16 |
| | | 5.00 | 1.48 | 0.04 | 98.52 | 19.18 | 0.13 | 64.85 | 0.16 | 15.98 | 0.12 | 37.41 |
| | | 2.00 | 1.60 | 0.04 | 98.40 | 19.21 | 0.11 | 65.42 | 0.14 | 15.38 | 0.10 | 41.51 |
| LEAP 5000 XS (355 nm) | None | 100.00 | 1.46 | 0.03 | 98.54 | 20.76 | 0.11 | 61.07 | 0.14 | 18.18 | 0.11 | 25.66 |
| | | 80.00 | 1.81 | 0.03 | 98.19 | 21.99 | 0.09 | 62.09 | 0.11 | 15.92 | 0.09 | 26.45 |
| | | 50.00 | 2.75 | 0.04 | 97.25 | 22.09 | 0.10 | 61.54 | 0.12 | 16.38 | 0.09 | 25.40 |
| | | 20.00 | 2.92 | 0.05 | 97.08 | 21.00 | 0.13 | 62.17 | 0.16 | 16.83 | 0.12 | 29.85 |
| | | 10.00 | 3.34 | 0.05 | 96.66 | 20.78 | 0.12 | 62.49 | 0.15 | 16.73 | 0.11 | 24.82 |
| | | 5.00 | 3.80 | 0.06 | 96.20 | 19.58 | 0.12 | 63.93 | 0.14 | 16.50 | 0.11 | 28.27 |
| | | 2.00 | 4.54 | 0.06 | 95.46 | 19.11 | 0.11 | 64.10 | 0.14 | 16.79 | 0.11 | 31.13 |
| LEAP 5000 XS (355 nm) | Cr 5nm | 100.00 | 2.04 | 0.05 | 97.96 | 20.50 | 0.14 | 67.71 | 0.17 | 11.79 | 0.12 | 48.86 |
| | | 80.00 | 1.58 | 0.03 | 98.42 | 21.57 | 0.09 | 66.24 | 0.10 | 12.19 | 0.07 | 38.86 |
| | | 50.00 | 1.40 | 0.03 | 98.60 | 21.57 | 0.12 | 64.41 | 0.14 | 14.02 | 0.10 | 36.17 |
| | | 20.00 | 1.33 | 0.03 | 98.67 | 21.28 | 0.11 | 63.30 | 0.13 | 15.43 | 0.10 | 34.17 |
| | | 10.00 | 1.84 | 0.03 | 98.16 | 20.13 | 0.10 | 63.07 | 0.12 | 16.80 | 0.09 | 27.56 |
| | | 5.00 | 2.06 | 0.05 | 97.94 | 19.35 | 0.13 | 63.64 | 0.16 | 17.01 | 0.13 | 29.59 |
| | | 2.00 | 2.54 | 0.04 | 97.46 | 18.66 | 0.11 | 64.34 | 0.13 | 16.99 | 0.10 | 32.56 |
| LEAP 6000 XR (260 nm) | Cr 5nm | 100.00 | 0.11 | 0.01 | 99.89 | 26.70 | 0.18 | 43.31 | 0.20 | 29.99 | 0.19 | 16.39 |
| | | 80.00 | 0.11 | 0.01 | 99.89 | 21.67 | 0.15 | 46.51 | 0.19 | 31.82 | 0.17 | 15.77 |
| | | 50.00 | 0.18 | 0.01 | 99.82 | 17.99 | 0.13 | 51.80 | 0.17 | 30.21 | 0.16 | 15.26 |
| | | 20.00 | 0.28 | 0.01 | 99.72 | 16.38 | 0.10 | 57.11 | 0.13 | 26.52 | 0.11 | 16.98 |
| | | 10.00 | 0.41 | 0.02 | 99.59 | 15.61 | 0.14 | 58.84 | 0.19 | 25.55 | 0.17 | 19.35 |
| | | 5.00 | 0.56 | 0.02 | 99.44 | 15.33 | 0.10 | 60.47 | 0.13 | 24.20 | 0.11 | 21.94 |
| | | 2.00 | 0.71 | 0.03 | 99.29 | 14.74 | 0.11 | 62.16 | 0.16 | 23.09 | 0.13 | 27.09 |
| Expected | - | - | 0.00 | - | - | 16.67 | - | 66.67 | - | 16.66 | - | - |

## 4) Discussion

Mass spectrum optimisation

The study of $CePO_4$ using APT presents a significant challenge, requiring a deeper understanding of field evaporation to achieve accurate results. Here, as a first approach, our aim was to identify the experimental parameters needed to optimize the mass spectrum, specifically by minimizing noise and maximising mass resolving power to prevent overlaps and enable better identification. Indeed, when analyzing complex materials (such as monazite), the presence of peak forests, the proximity of certain species and the overlaps in the mass spectrum remain one of the main limitations of atom probe tomography. This is why we believe that optimizing the mass spectrum (particularly its resolving power and noise level) could help reduce some of the losses associated with its processing. To this end, we first compared two laser wavelengths, 260 and 355 nm, on two straight atom probes: the PAP and the LEAP 5000 XS. As mentioned in the theoretical section, using a shorter wavelength increases the energy of the photons : $E_{260}$ = 4.77 eV and $E_{355}$ = 3.5 eV.

In the literature, the band gap value of $CePO_4$ is reported to range between 4.3 and 5.28 eV, depending on the source (Adelstein et al. 2011; Kirubanithy et al. 2015). This value is lowered by the electric field and charge accumulation at the surface of the sample, causing the energy of incident photons to exceed the band gap energy. This significantly enhances laser absorption by the specimen. For illumination at a wavelength of 260 nm, this should result in thermal activation of only the outer surface of the specimen and rapid cooling, and should improve peak resolution. Experimentally, a large difference in background noise is observed between the two instruments, making the comparison challenging (Figure 4). However, the slightly higher resolution of the $O_2$ peak at 50% for the experiment using a laser wavelength of 260 nm (PAP instrument) could suggest a greater contribution of surface absorption to the thermal component of evaporation. Meanwhile, the noise level and the shape of the peaks (with more pronounced thermal tails) indicate a higher heating of the bulk of the sample, possibly caused by a larger laser spot size compared to the other instruments or poor laser contrast, leading to the absorption of a significant amount of residual energy by the sample between pulses. Another hypothesis is that the noise could originate from the laser pulse itself, where highly energetic photons generate electrons in the analysis chamber, causing significant background noise on the detector. This hypothesis is supported by the very low background noise observed in experiments conducted with the LEAP 6000 XR, which features technology absent in the PAP instrument. The PAP instrument is optimized to improve the collection angle of photons that could be emitted from the sample. To achieve this ability, the space between the tip and the detector was let opened (Figure 1.A). Conversely, the detector of the LEAP 6000 XR is encapsulated to avoid any noise induced by wandering ions that could be emitted outside the tip apex. In addition, the LEAP 6000 XR is equipped with technology that allows the detection to be momentarily paused during the laser pulse, effectively preventing such interferences induced by direct impact of UV photons on the detector.

We assume that the thermal absorption profile on the sample surface increases the thermal contribution to ion evaporation. The experiments suggest a difference in the shape and resolution of the peaks depending on the LPE, with a more pronounced effect observed for a shorter wavelength (260 nm). This supports the hypothesis that a shorter wavelength enhances the rapid component of thermally assisted evaporation (Figure 4).

As demonstrated by Silaeva et al. (Silaeva et al. 2014), the electric field applied to an insulating material alters the physical characteristics of its outer surface, reducing the band gap and creating a metallic surface state that enhances laser absorption. The wavelength should have a lower effect on the absorption at high electric fields where the band gap reduction is strong enough for the surface to behave like a metal. However, at lower fields, the difference in response becomes much more pronounced. Similarly, adding a CrO layer to the sample surface allows both wavelengths to be absorbed with almost the same efficiency. Moreover, this metallic coating homogenizes heat distribution across the sample surface, improving mass resolution and reducing the variance in flight times for ionic species (Kwak et al. 2017). The effect of the coating on laser absorption is more pronounced for analyses performed on the LEAP 5000 XS (355 nm). As shown in the spectra presented in Figure 5, the coated sample analysed with the LEAP 5000 XS exhibits significantly lower noise levels and improved peak resolution, both attributed to the enhanced surface absorption provided by the coating. In contrast, the samples analysed on the PAP instrument (260 nm) show much more similar spectra, with the coated sample displaying only slightly lower noise levels, likely due to the conductivity of the coating. However, there is almost no difference in peak resolution.

As mentioned before, analyses conducted on the LEAP 6000 XR revealed significantly lower noise levels compared to other instruments (Figure 6). This, combined with the improved resolution provided by the reflectron and the laser wavelength (260 nm), allowed for the detection of additional peaks that were previously hidden by background interference or thermal tails. This enables the identification of extra molecular species or minor isotopes of certain elements for the four $Ce^{3+}$ peaks at 45.3, 45.9, 46.6 and 47.3 Da easier to measure on the LEAP 6000 XR spectrum. The mass spectrum thus becomes easier to analyse, and the detection of additional minor isotopes can be used for deconvoluting overlapped peaks, potentially leading to more accurate compositional measurements. Nevertheless, the compositional measurements obtained with this instrument and presented in Table 2 show larger differences than those from all the other instruments, particularly at high LPE values. This suggests that the compositional biases observed in monazite during APT analyses do not arise solely from interpretation errors or a lack of mass spectrum optimization, but also from evaporation mechanisms leading to preferential losses of certain elements.

Composition measurement

To correlate the compositional measurements in $CePO_4$ with the optimization of the mass spectrum, the atomic composition measured in each of the experiments described above is summarized in Table 2. Many studies have shown that measuring composition using APT in insulating materials containing oxygen is challenging and subject to significant uncertainties coming from multiple sources (Cappelli et al. 2021; Santhanagopalan et al. 2015; Sen et al. 2021; Torkornoo et al. 2024; Jakob et al. 2024) but primarily due to oxygen loss, mainly at high LPE. Devaraj et al. (Devaraj et al. 2013) proposed that the increase in oxygen deficiency with rising laser energy may result from the formation of neutral oxygen molecules at lower evaporation fields. They argued that reducing the UV laser energy raises the electric field at the specimen (due to the higher standing voltage required to sustain the same evaporation rate), which increases the probability of ionizing neutral molecules that desorb. However, this hypothesis cannot be directly confirmed by APT, as neutral molecule loss is undetectable (Diercks et al. 2013). Morris et al. (Morris et al. 2024) propose a similar model for the evaporation of BaSrTiO layers, where the concentration of non-evaporated oxygen in molecular $TiO^{2+}$ ions on the sample surface increases until thermally desorbed neutral $O_2$ molecules form.

The compositions measured in this study revealed systematic errors in the three main elements, Ce, P, and O, which appear to originate from multiple sources, primarily linked to the applied pulse energy (LPE) and, consequently, to the electric field. In most cases, the amounts of P and Ce decrease as the electric field increases, while the O composition rises. This effect is particularly noticeable in the LEAP 6000 XR experiment, where the noise level is very low. Some studies have shown that consistent elemental ratios can be obtained when oxygen or nitrogen are excluded from the composition calculations (Morris et al. 2018). In our case, we sought to better understand the mechanisms responsible for elemental losses in order to improve the overall accuracy of composition measurements. The results presented in Table 2 show that the Ce and P concentrations vary with the analysis conditions, seemingly in correlation with oxygen, although no clear trend can be established. Since Ce and P should theoretically be present in equal proportions, we examined the Ce/P ratio. This ratio is closer to the expected value under high-field conditions, where the oxygen deficit is lower, suggesting that oxygen loss may also influence the measured quantities of other elements. However, the ratio remains much lower for analyses performed with the LEAP 6000 XR, due to a higher apparent Ce concentration compared with other instruments. We argue that this discrepancy most likely arises from the reduced ability of the LEAP 6000 XR to detect multiple events, although we cannot confirm this with certainty. The detailed Ce/P ratio data and corresponding plots are available in the supplementary material.

Therefore, as a first step, it is essential to investigate the variations in the electric field as a function of the instrument and laser energy. However, evaluating the electric field using traditional methods (Kingham 1982) can be particularly complex in the case of $CePO_4$, as Ce is the only element present in two distinct charge states: $Ce^{3+}$ (46.6 Da) and $Ce^{2+}$ (70 Da) in its elemental form (outside molecular ions). This challenge is further compounded by the proximity of the $Ce^{3+}$ peak to the $PO^+$ peak (46.9 Da), making differentiation difficult when resolution is low, and by the overlap of the $Ce^{2+}$ peak with the tail of the $PO_2^+$ peak (62.9 Da), introducing uncertainty in its intensity.

Nevertheless, a qualitative analysis remains possible and results are presented in Figure 8.A For experiments without a Cr coating, the electric field appears to change rather abruptly between 80 and 100 pJ. The measured compositions, which also deviate from overall trends, suggest that the results are strongly biased by the higher noise levels and increased thermal tails observed at these LPE values.

An alternative approach, based on the evolution of VDC with varying LPE, was used to qualitatively assess the relative variation of the electric field between instruments (Figure 8.B). In this method, we start from an arbitrary initial field value and observe the change in VDC as the laser energy increases. Since the electric field is directly related to VDC (Vella et Houard 2016), the relative variation of the field can be qualitatively inferred. The comparison of these curves between the LEAP 6000 XR and the other instruments shows that, for similar laser energy ranges, the electric field does not evolve in the same way. This indicates a distinct thermal contribution to evaporation depending on the laser wavelength. The difference with the PAP, which uses the same wavelength as the LEAP 6000 XR, could instead be attributed to a different laser focusing (spot size) or to differences in the shape of the samples used for each experiment. Additionally, the higher presence of hydrogen in the LEAP 5000 XS and the PAP instrument could lead to a reduction in the field. The combination of these phenomena could explain why the field variation is less significant for the LEAP 6000 XR.

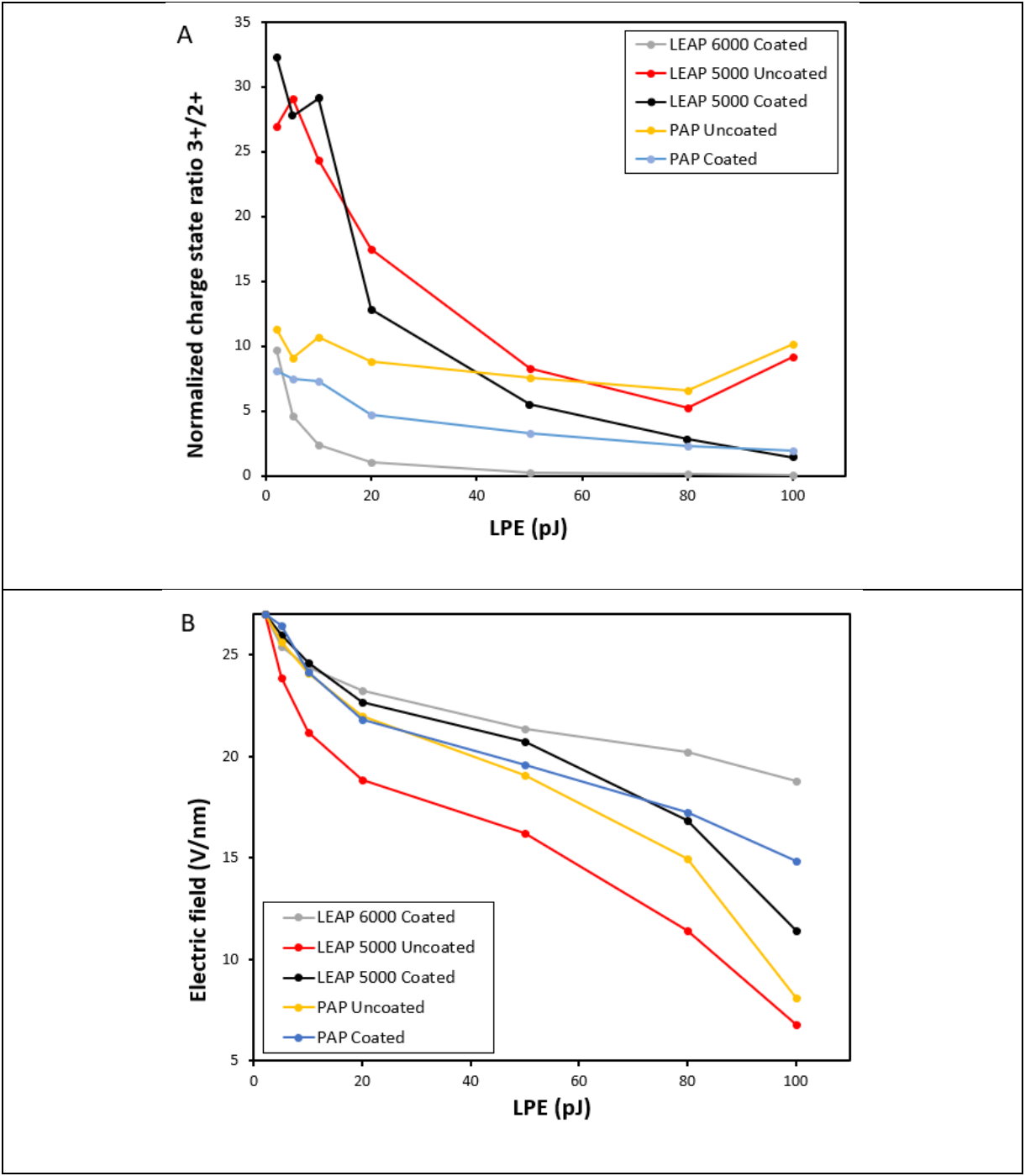

*Figure 8: A) Evolution of the Ce3+ on Ce2+ ratio as a function of LPE for $CePO_4$ analyzed with different instruments. B) Evolution of the relative electric field as a function of LPE for $CePO_4$ analyzed with different instruments.*

Some studies (Santhanagopalan et al. 2015; Jakob et al. 2024) suggest that oxygen loss results from molecular dissociation during flight, forming neutral $O_2$ molecules that remain undetected. In our case, most of the oxygen detected in the mass spectrum appears in the $O_2$ peak (32 Da). Figures 9.A & B present the measured $O_2$ composition and the percentage of multi-hit events as a function of LPE. Despite the detector's higher sensitivity to multiple events (Costa et al. 2012), the PAP instrument does not significantly improve the measurement.

Moreover, the number of detected multi-hit events remains consistently low across all instruments (on the $O_2$ peak) and shows only minor variations with changing field strength—typically around 1% in most cases. This suggests that although pile-up effects may contribute to compositional losses, their overall impact appears secondary compared to the significant oxygen variations observed, which can reach several percent. Particular attention is given to the LEAP 6000 XR, which exhibits the largest compositional deviation in $O_2$ despite recording the lowest and least variable rate of multi-hit events. We propose that this discrepancy arises from the specific configuration of the instrument, where the use of a reflectron focuses ions from a wide angular range onto a comparatively small detector, thereby increasing the likelihood of pile-up losses (Ndiaye et al. 2023). Additionally, the presence of the grid associated with the reflectron may further amplify this effect by scattering ions and reducing detection efficiency, as discussed by Thuvander et al. (2013).

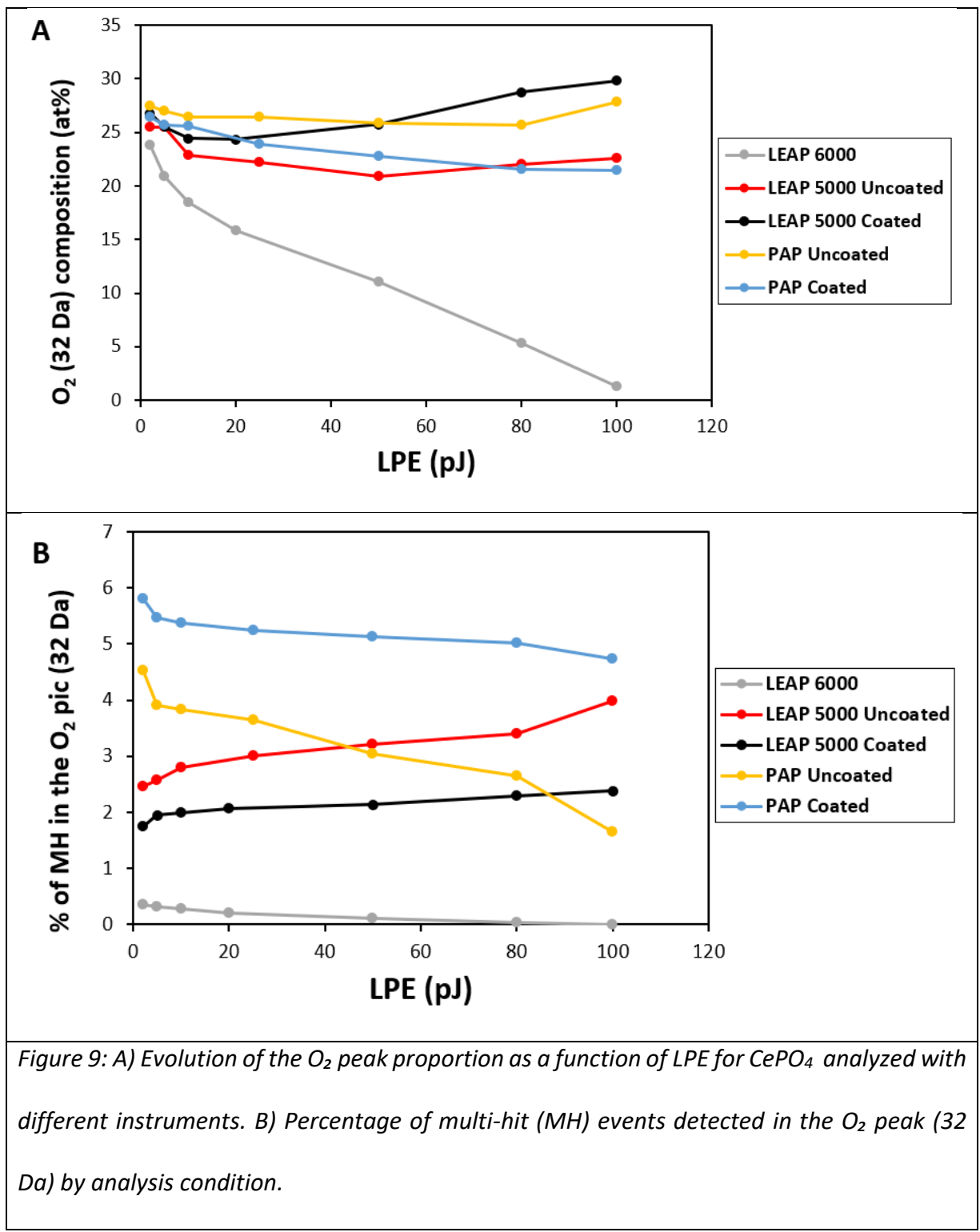

Figure 9: A) Evolution of the $O_2$ peak proportion as a function of LPE for $CePO_4$ analyzed with different instruments. B) Percentage of multi-hit (MH) events detected in the $O_2$ peak (32 Da) by analysis condition.

The analysis of correlation diagrams (Saxey 2011), an example of which is shown in Figure 10, reveals dissociation lines but no clear evidence of neutral formation as reported by Gault et al. (2016). Moreover, the number of ions involved in these dissociations represents less than 0.2% of the total number of atoms in the experiment. This indicates that post-evaporation dissociation alone cannot account for the substantial oxygen loss observed. However, it remains possible that dissociations occurring very close to the specimen surface contribute to composition biases. In such cases, these events are effectively equivalent to neutral emission, as the dissociated species are not accelerated and thus not detected, potentially leading to local pile-up effects and further information loss (Gault et al. 2016; Zanuttini et al. 2017).

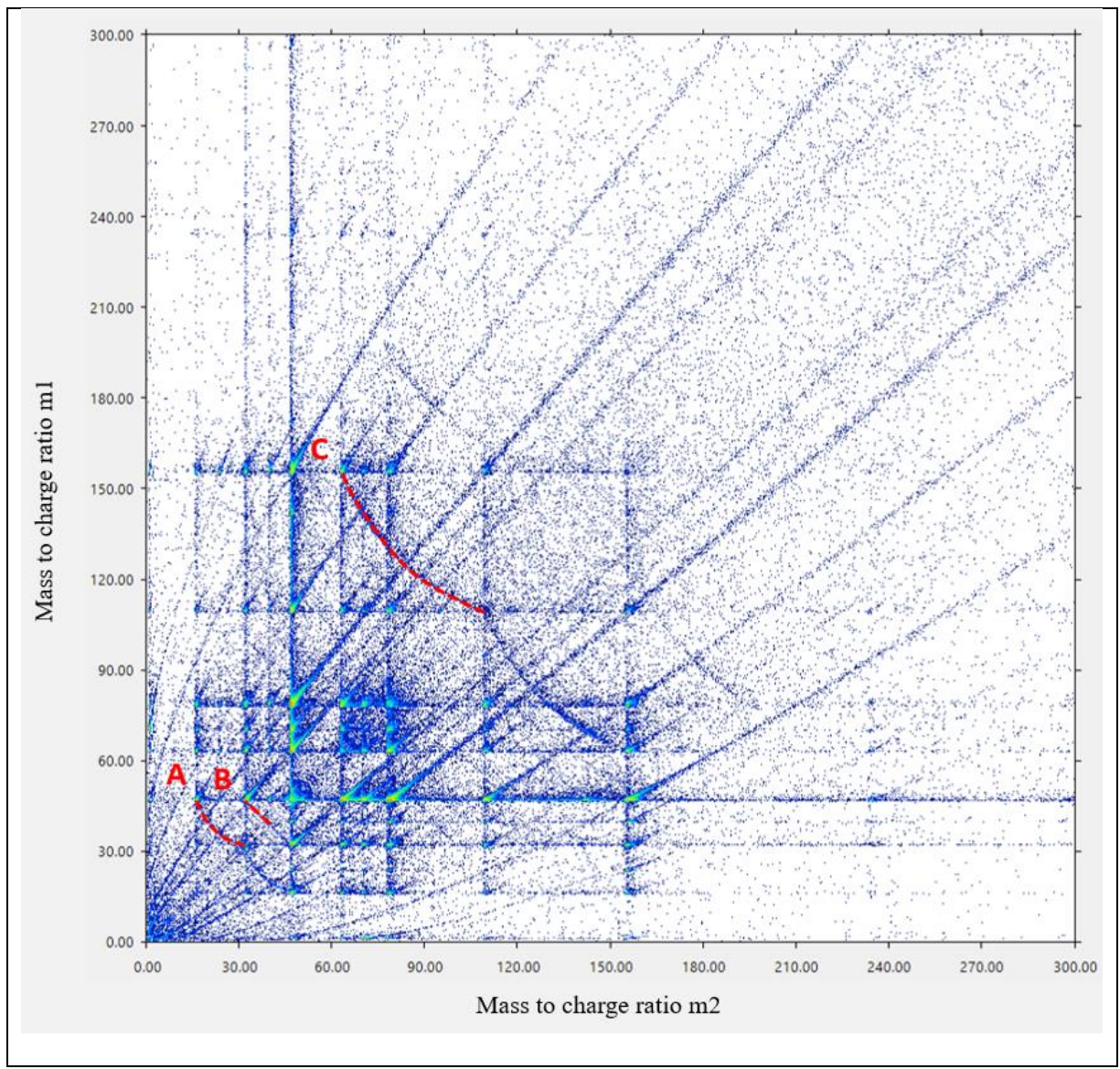

*Figure 10: Correlation diagram of a CePO$_4$ sample obtained with the PAP instrument. Principal dissociation lines are noted as : A) PO$_2$$^{2+}$ → O$^+$ + PO$^+$; B) PO$_3$$^+$ → O$_2$$^+$ + PO$^+$; C) CePO$_3$$^{2+}$ → CeO$^+$ + PO$_2$$^+$.*

The apparent loss in oxygen concentration in the analysis of CePO$_4$ is strongly correlated with the surface electric field. Although a systematic bias is always present, the composition approaches the correct value at high field/low energy laser conditions. This observation is in line with previous studies of oxides and nitrides materials (Santhanagopalan et al. 2015; Markus, Karahka and Kreuzer 2015; Morris et al. 2024; Valderrama et al. 2015; Jakob, Fazi, et Thuvander 2024; Schiester et al. 2024). In these experimental cases, most of authors claimed a deficiency induced by the possibility of the surface to directly emit neutral O$_2$ or N$_2$ in conditions where the field was not sufficient to induce field ionisation. As stated by Gault et al. (Gault et al. 2016), no direct experimental or theoretical evidence of this neutral was demonstrated because if neutrals are emitted, the atoms were not accelerated by the present electric field and detected by the ion detector. On a zinc oxide (ZnO) surface, Xia et al. simulated the surface stability under a positive electric field ranging from 25 to 30 V/nm (Karahka et al. 2015). Above 15 V/nm, the surface undergoes reorganization due to the electrostatic field effect, and beyond 25 V/nm, the geometry optimization no longer converges. The surface rapidly emits metallic species (in the form of Zn$^{2+}$), while oxygen atoms rearrange quickly to form O$_2$ molecules, which become instantaneously ionized and leave the surface as O$_2$$^+$. We may note that oxygens atoms are never first neighbours in the initial crystal structure, indicating a spontaneous reorganisation with no barrier in density functional theory (DFT). This reorganization subsequently leads to the emission of other, more complex molecules, which can eventually dissociate and be ejected in their cationic forms. This process continues as long as the electric field is applied. This theoretical mechanism appears to correspond to experimental observations in various oxides and nitrides. While many authors seem to identify the direct emission of neutral molecules from the surface, one may question the likelihood of such an event occurring on our sample, which is subjected to an electric field exceeding 20 V/nm.

Indeed, two physical phenomena oppose this mechanism. First, under these field strengths, if neutral particle emission were to occur, the particle would be strongly polarized by the electric field. B. Gault et al. indicated that under the influence of the electric field, and due to the atomic polarizability of $O_2$, the binding energy $E_p$ associated with surface polarization is on the order of 0.2 to 0.45 eV under a field between 20 and 30 V/nm ($\alpha \sim 1.6 \times 10^{-40} F/m^2, E_p = \frac{1}{2}\alpha F^2$). A molecule desorbed due to thermal excitation (a few hundred K, 10 to 100 meV) possesses a kinetic energy far below this threshold and should theoretically remain attached to the surface (the binding force follows the electric field gradient, which extends over long distances).

Furthermore, density functional theory (DFT) models consistently predict the emission of positively charged particles. This is fundamentally linked to the limitations of DFT, which neglects reaction kinetics and assumes the electronic ground state at each time step during relaxation. In reality, it should be necessary to evaluate the probability of ionization throughout the entire process of molecule or atom expulsion. This kind of calculation was done in the process of field ionization or post field ionisation, in the context of a single atom or molecule escaping the surface. To evaluate this probability, we make use of the post-ionisation theory, developed from the ionisation theory described in the early ages of field evaporation to explain the emission mechanism. This approach calculates the integrated probability of ionisation of a particle traveling from the specimen surface to the infinite using Schrödinger equation in simple model potential of the atom under the presence of a surface electrostatic field. For a single particle of charge leaving a planar surface submitted to a field F, the probability of ionisation or post ionisation P(n,F) follows equation

$$P(n, F) = 1 - exp\left[-\int_{zc}^{\infty} \frac{R(z,n,F)}{v(z,n,F)} dz\right] \quad (3)$$

Where $v(z,n,F)$ is the velocity of the particle, $R(z,n,F)$ the rate constant of ionization, n the degree of ionization, and z the distance to the surface. $z_c$ is the critical distance of ionisation, which is the minimal distance for which an electron can tunnel from the particle to the surface (in the range of 2 to 5 Angström to the surface). R is found from the electron probability flux of the outermost orbital through a surface *S* perpendicular to the field direction

$$R = \int_S |\psi|^2 u_e dS \quad (4)$$

$\psi$ is the electron wavefunction that is calculated using the solution of Schrödinger equation in the JWKB approximation and $u_e$ the electron velocity normal to the surface.

For an atom in free space, an approximation of R is

$$R \sim \omega \times exp\left(-\frac{b}{F}\right) \quad (5)$$

b is a constant depending mostly of the ionisation energy, and of the shape of the energy barrier. In the approximation of a triangular barrier $b = \frac{2}{3}\frac{\sqrt{32\pi^2 m_e}}{eh} I^{3/2}$. ω is a parameter weakly dependant of F. Considering equation (3) and (5), P(n,F) can be approximated by equation (Lefebvre-Ulrikson 2016) :

$$P(F) = 1 - exp\left[-a \times \exp\left(-\frac{b}{F}\right)\right] \qquad (6)$$

With a and b, two ad-hoc constants.

In the context of post-ionisation of an ion leaving the surface, the influence of the surface in high electric field on the energy barrier must be considered. It was estimated carefully by Kingham et al., and documented in several publications (Kingham 1982; Ernst 1979; Tegg et al. 2024; Cuduvally et al. 2022), but the same trend of P with F can be found. We may note that the velocity $v$ of the ion need to be carefully estimated in equation 3, and it was done assuming the Muller model of field evaporation, with an initial null velocity of the ion. Note that the model was also slightly adjusted to experimental observations in the post-ionisation of rhodium.

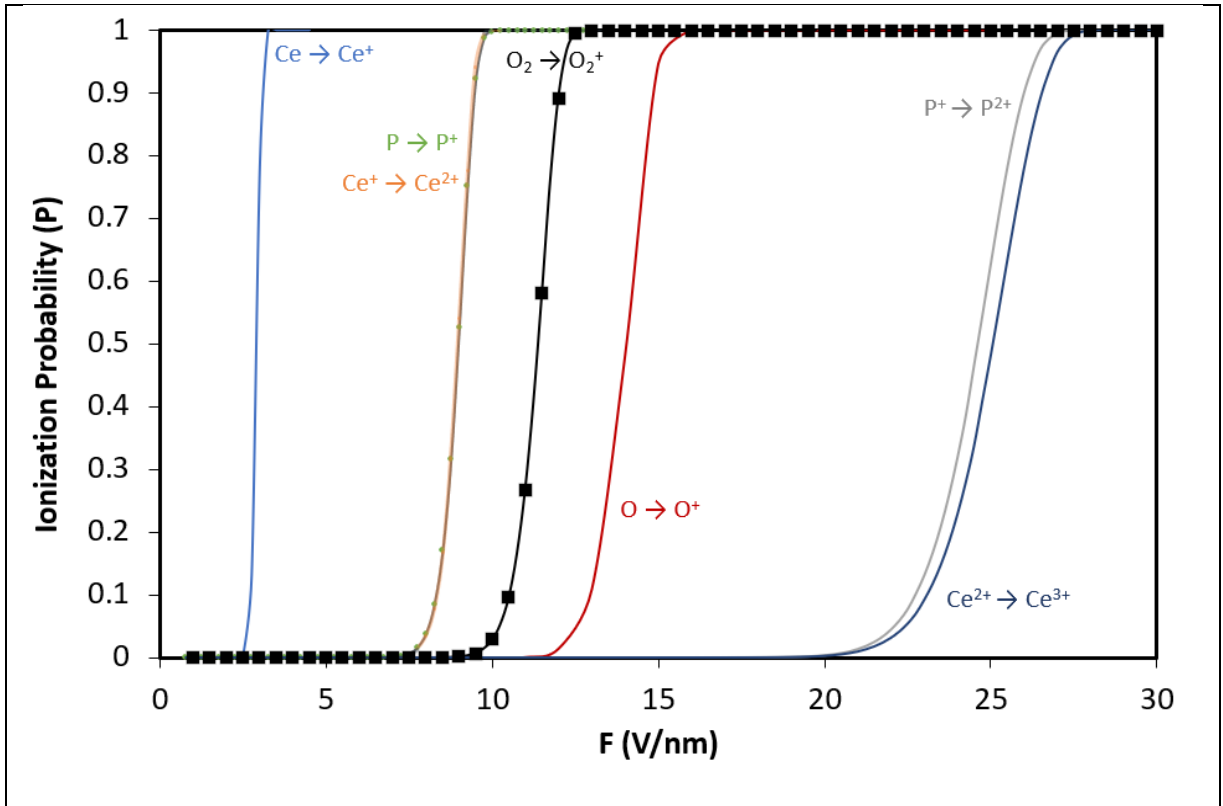

Figure 11: Ionisation or post-ionisation probabilities calculated using the Kingham's model as a function of the surface electric field. Note that P → P⁺ transition is close to be *superimposed to Ce$^+$ → Ce$^{2+}$ curve. The O$_2$ → O$_2^+$ curve was fitted (less square method) using the simple model P=1-exp(-a.exp(-b/F)), with a=6.2x10$^9$ and b=260.8 (black squares).*

We used this model to estimate the ionisation and post ionisation probability of O$_2$ in Figure 11. We may note that this model gives an absolute estimation of the threshold fields with an uncertainty of about 15% as stated by Kingham. The probability of ionisation at high filed is found to be stronger than 15 V/nm for all the elements. Note the superimposition of the (P → P$^+$) probability curve to the (Ce$^+$ → Ce$^{++}$) probability curve since the second ionisation energy if the Ce (10.96 eV) is close to the first ionisation energy of P (10.49 eV), showing the strong influence of this parameter on the threshold fields of ionisation. Considering experimental Ce$^{3+}$/Ce$^{2+}$ measurements, the electric field in experiments is estimated in the range 20 to 30 V/nm, normally strong enough to fully ionise all atomic or molecular species. We fitted the (O$_2$ → O$_2^+$) probability curve with the model of equation (6). The agreement is surprisingly good, giving an adjustment of the *b* coefficient of 260.8 (F in V/nm) close to the expected coefficient of 286.5 assuming I=12.07 eV. Considering the shape of this curve, this model predicts that a desorbed molecule could not leave the surface without being ionised.

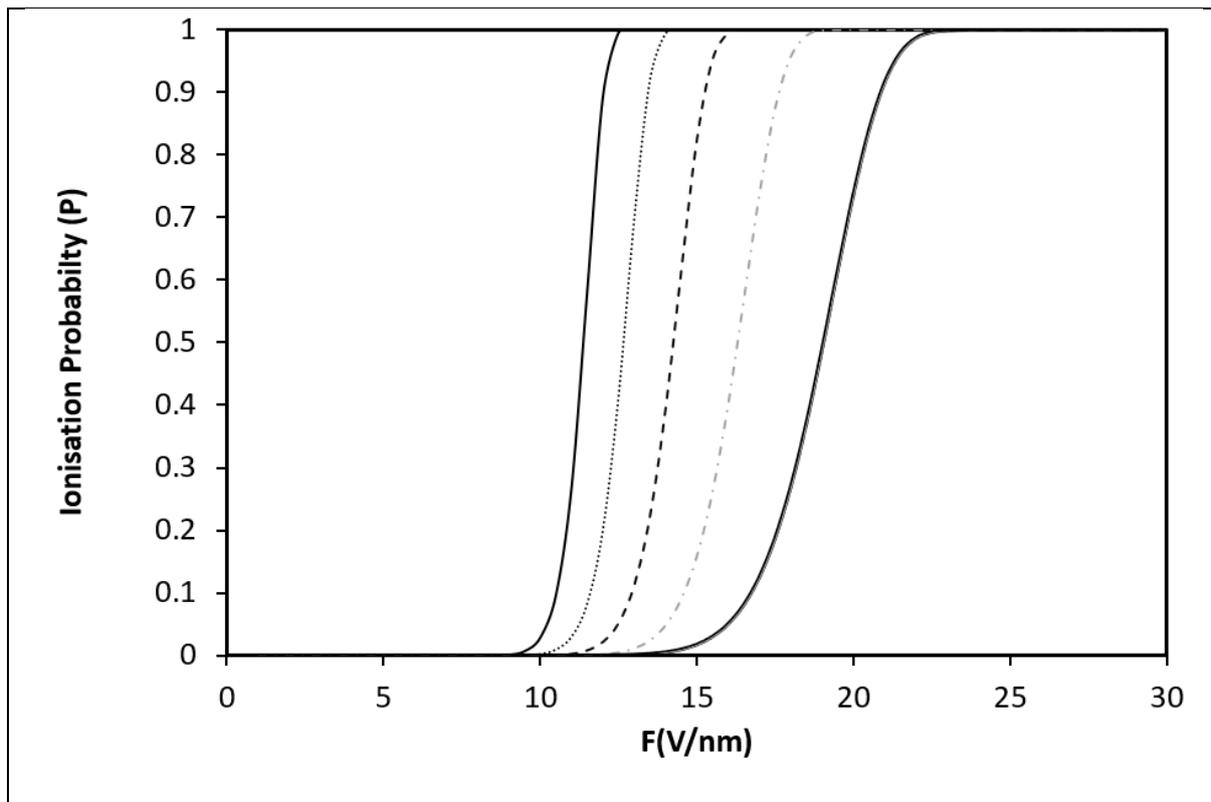

*Figure 12: Ionisation probabilities curves assuming different initial velocity of the emitted particle starting from the Kingham model (thin black line). The velocity of the particle is multiplied by 10 between each curve.*

The only parameter that could explain the desorption of neutral special without field induced ionisation is the presence of a significant initial velocity of the emitted particle. Indeed, the Kingham model predicts that the zone of ionisation calculated in the integral of equation 3 is extremely thin (Haydock et Kingham 1981). Most of the ionisation takes place very close to the surface (at the critical distance of ionisation). As stated by Gomer (Gomer 1994), for an atom leaving the surface, ionisation and evaporation takes place almost at the same place. Ionisation probability is thus limited in equation 3 by the time of residence of the particle at this position which is inversely proportional to the velocity of the atom or the molecule. In Figure 12, we recalculate the probability, by changing the parameter *a* including this velocity.

We may introduce a new constant parameter a', introducing the initial velocity. As a result, $a' = a \frac{v0}{v1}$ with $v_1$ the velocity of the emitted particle and $v_0$ the velocity of the particle in the Kingham model. As observed in Figure 12, a shift to higher threshold field to ionise emitted particle is predicted simply assuming a faster initial velocity. A significant increase in velocity is possible if we assume that the molecule leaves the surface in an excited state, with kinetic energy of more than 1 eV. In addition, this process is compatible with the required energy to be liberated from the polarisation effect described previously (The kinetic energy must be higher than 0.2-0.4 eV).

We may note that molecules in excited states in APT were observed concurrently in some other studies during the process of field evaporation (Vurpillot et al. 2018; Zanuttini et al. 2017). Indeed, some dissociation channels observed experimentally were associated to the presence of strongly excited internal states in $SiO_2$ or GaN, normally not existing at moderate temperature (<1000K). Shifts in the energy curves of electrons could reach several eV, indicating that the process of field evaporation significantly populates these states. We postulate that the atomic surface reconstruction following the emission of metallic species, as modelled by Kreuzer et al. could produce excited molecules. It could be likely for oxygen atoms, that are not linked together in the original structure of the monazite (oxygens atoms are bounded to the rare earth element and to the phosphorous atom). The FIM experiments carried out in this work, and presented in Figure 7.A & B, show, through the absence of poles and visible crystallographic structures on the surface, significant surface reconstruction phenomena, probably largely promoted by the thermal energy delivered by the laser pulses during the atom probe experiments. When the reconstruction takes places, the formation of the oxygen molecule, strongly bounded, is exothermic and liberate locally some energy. A part of this energy could be transformed to kinetic energy during the process of liberation. If the velocity is high enough, the residence time will be short, and the ionisation of the particle could be impeded. Note that presence of significant kinetic energy spreads was also observed when analysing the energy distribution of emitted ions in laser assisted evaporation (Sévelin-Radiguet et al. 2015).

The increase in the field is also correlated with a rise in background noise in mass spectra, which may indicate preferential evaporation. In other words, elements with lower evaporation fields could be evaporated between pulses, preventing time-of-flight measurement and thus their identification. In our case, cerium, has a relatively low evaporation field. Moreover, the measured compositions show a decrease in cerium concentration directly linked to the field increase, reinforcing the hypothesis of preferential evaporation as sources of biases.

It thus appears that the composition biases observed in $CePO_4$ result from a combination of various physical phenomena occurring during evaporation, leading to selective losses, especially at high electric fields.

## 5) Conclusion

This study demonstrates the potential of Atom Probe Tomography (APT) for nanoscale analysis of insulating geological materials such as $CePO_4$, while also highlighting the challenges associated with accurate compositional measurements. The influence of experimental parameters, including laser wavelength, laser pulse energy (LPE), and the application of coatings, was systematically investigated to optimize mass spectrum resolution and minimize background noise.

The use of shorter laser wavelengths (~260 nm) clearly improved the resolution of mass spectrum peaks, particularly the major ones, and significantly reduced thermal tails. This enhancement was most notable when employing advanced configurations such as the reflectron-equipped LEAP 6000 XR, which achieved superior mass resolution compared to straight-flight instruments. The combination of a deep UV wavelength and reflectron technology enabled the identification of additional minor peaks, facilitating the deconvolution of overlapping spectra and improving compositional accuracy. However, the smaller detector size in this instrument, compared to others, led to considerable information loss, likely due to pile-up effects.

Applying thin CrO coatings to the sample needles proved effective in further reducing noise levels and enhancing thermal conductivity, particularly for instruments operating at higher LPEs. These coatings improved both heat dissipation and mass resolution, with the most significant effects observed for the LEAP 5000 XS. Nonetheless, the benefits of coatings were less pronounced for the PAP instrument, suggesting that their effectiveness may depend on the laser wavelength used.

Despite these advances, systematic biases in compositional measurements were consistently observed. These discrepancies are linked to mechanisms such as neutral molecule desorption for oxygen, preferential evaporation of metallic elements, and the previously mentioned pile-up effects. Additionally, increased background noise and elevated hydrogen content at higher electric fields further complicated compositional quantification, highlighting the need for careful optimization of experimental conditions.

In conclusion, this study emphasizes the critical role of optimizing experimental parameters in APT to enhance data quality and reliability when analyzing insulating materials like $CePO_4$. While significant progress has been achieved, challenges remain in mitigating oxygen loss and improving quantification accuracy. Future work should focus on developing advanced analytical models and experimental protocols to overcome these limitations, thus enabling broader applications of APT in geological and cosmochemical research.

The results presented here show that, for the analysis of phosphates by APT, it is advisable to use, whenever possible, a wavelength in the deep UV range (~260 nm) combined with a reflectron, in order to maximize mass spectrum resolution and facilitate interpretation. It is also recommended to evaporate atoms under high electric fields (low laser pulse energy, ~1–2 pJ) to achieve more accurate compositional measurements, although the mechanisms responsible for these improvements are still under discussion.

Acknowledgments

We would like to express our heartfelt thanks to J.M. Montel for generously providing the samples used in this study. We also thank Fabien Cuvilly for his support in sample preparation using SEM and FIB, as well as Celia Castro for her assistance with TEM. The authors are also grateful to Frederic Danoix, Angela Vella, Gérald Da-Costa, and Sylvain Nully for their valuable feedback, discussions, and support.

# References

Adelstein, Nicole, B. Simon Mun, Hannah L. Ray, Philip N. Ross, Jeffrey B. Neaton, et Lutgard C. De Jonghe. 2011. « Structure and Electronic Properties of Cerium Orthophosphate: Theory and Experiment ». *Physical Review B* 83 (20): 205104. https://doi.org/10.1103/PhysRevB.83.205104.

Arnoldi, L., E.P. Silaeva, F. Vurpillot, et al. 2015. « Role of the Resistivity of Insulating Field Emitters on the Energy of Field-Ionised and Field-Evaporated Atoms ». *Ultramicroscopy* 159 (décembre): 139-46. https://doi.org/10.1016/j.ultramic.2014.11.018.

Arnoldi, L, P Silaeva, A Gaillard, et et al. 2014. « Energy deficit of pulsed-laser field-ionized and field-emitted ions from non-metallic nano-tips ». *Journal of Applied Physics*.

Blavette, Didier, Talaat Al Kassab, Emanuel Cadel, et al. 2008. « Laser-Assisted Atom Probe Tomography and Nanosciences ». *International Journal of Materials Research* 99 (5): 454-60. https://doi.org/10.3139/146.101672.

Boatner, Lynn A. 2002. « Synthesis, Structure, and Properties of Monazite, Pretulite, and Xenotime ». *Reviews in Mineralogy and Geochemistry* 48 (1): 87-121. https://doi.org/10.2138/rmg.2002.48.4.

Bogdanowicz, J. 2018. *Laser-Assisted Atom Probe Tomography of Semiconductors: The Impact of the Focused-Ion Beam Specimen Preparation*.

Bunton, Joseph H., Jesse D. Olson, Daniel R. Lenz, et Thomas F. Kelly. 2007. « Advances in Pulsed-Laser Atom Probe: Instrument and Specimen Design for Optimum Performance ». *Microscopy and Microanalysis* 13 (6): 418-27. https://doi.org/10.1017/S1431927607070869.

Cappelli, Chiara, Sandi Smart, Harold Stowell, et Alberto Pérez-Huerta. 2021. « Exploring Biases in Atom Probe Tomography Compositional Analysis of Minerals ». *Geostandards and Geoanalytical Research* 45 (3): 457-76. https://doi.org/10.1111/ggr.12395.

Clavier, Nicolas, Renaud Podor, et Nicolas Dacheux. 2011. « Crystal Chemistry of the Monazite Structure ». *Journal of the European Ceramic Society* 31 (6): 941-76. https://doi.org/10.1016/j.jeurceramsoc.2010.12.019.

Costa, G. Da, H. Wang, S. Duguay, A. Bostel, D. Blavette, et B. Deconihout. 2012. « Advance in Multi-Hit Detection and Quantization in Atom Probe Tomography ». *Review of Scientific Instruments* 83 (12): 123709. https://doi.org/10.1063/1.4770120.

Cuduvally, Ramya, Richard J. H. Morris, Giel Oosterbos, et al. 2022. « Post-Field Ionization of Si Clusters in Atom Probe Tomography: A Joint Theoretical and Experimental Study ». *Journal of Applied Physics* 132 (7): 074901. https://doi.org/10.1063/5.0106692.

Devaraj, A., R. Colby, W. P. Hess, D. E. Perea, et S. Thevuthasan. 2013. « Role of Photoexcitation and Field Ionization in the Measurement of Accurate Oxide Stoichiometry by Laser-Assisted Atom Probe Tomography ». *The Journal of Physical Chemistry Letters* 4 (6): 993-98. https://doi.org/10.1021/jz400015h.

Diercks, David R., Brian P. Gorman, Rita Kirchhofer, Norman Sanford, Kris Bertness, et Matt Brubaker. 2013. « Atom Probe Tomography Evaporation Behavior of C-Axis GaN Nanowires: Crystallographic, Stoichiometric, and Detection Efficiency Aspects ». *Journal of Applied Physics* 114 (18): 184903. https://doi.org/10.1063/1.4830023.

Ernst, Lajos. 1979. « On the Field Penetration into Semiconductors in the Field Ion Microscope ». *Surface Science* 85 (2): 302-8. https://doi.org/10.1016/0039-6028(79)90253-X.

Exertier, F., A. La Fontaine, C. Corcoran, et al. 2018. « Atom Probe Tomography Analysis of the Reference Zircon Gj-1: An Interlaboratory Study ». *Chemical Geology* 495 (septembre): 27-35. https://doi.org/10.1016/j.chemgeo.2018.07.031.

Fougerouse, D., S.M. Reddy, D.W. Saxey, et al. 2018. « Nanoscale Distribution of Pb in Monazite Revealed by Atom Probe Microscopy ». *Chemical Geology* 479 (février): 251-58. https://doi.org/10.1016/j.chemgeo.2018.01.020.

Fougerouse, Denis, Aaron J. Cavosie, Timmons Erickson, et al. 2021. « A New Method for Dating Impact Events – Thermal Dependency on Nanoscale Pb Mobility in Monazite Shock Twins ». *Geochimica et Cosmochimica Acta* 314 (décembre): 381-96. https://doi.org/10.1016/j.gca.2021.08.025.

Fougerouse, Denis, Christopher L. Kirkland, David W. Saxey, et al. 2020. « Nanoscale Isotopic Dating of Monazite ». *Geostandards and Geoanalytical Research* 44 (4): 637-52. https://doi.org/10.1111/ggr.12340.

Gardés, Emmanuel, Olivier Jaoul, Jean-Marc Montel, Anne-Magali Seydoux-Guillaume, et Richard Wirth. 2006. « Pb Diffusion in Monazite: An Experimental Study of Interdiffusion ». *Geochimica et Cosmochimica Acta* 70 (9): 2325-36. https://doi.org/10.1016/j.gca.2006.01.018.

Gault, Baptiste, David W Saxey, Michael W Ashton, et al. 2016. « Behavior of Molecules and Molecular Ions near a Field Emitter ». *New Journal of Physics* 18 (3): 033031. https://doi.org/10.1088/1367-2630/18/3/033031.

Gomer, Robert. 1994. « Field Emission, Field Ionization, and Field Desorption ». *Surface Science* 299-300 (janvier): 129-52. https://doi.org/10.1016/0039-6028(94)90651-3.

Gopon, Phillip, James O Douglas, Frederick Meisenkothen, Jaspreet Singh, Andrew J London, et Michael P Moody. 2022. « Atom Probe Tomography for Isotopic Analysis: Development of the 34S/32S System in Sulfides ». *Microscopy and Microanalysis* 28 (4): 1127-40. https://doi.org/10.1017/S1431927621013568.

Haley, D., P. Choi, et D. Raabe. 2015. « Guided Mass Spectrum Labelling in Atom Probe Tomography ». *Ultramicroscopy* 159 (décembre): 338-45. https://doi.org/10.1016/j.ultramic.2015.03.005.

Harrison, T. M., E. J. Catlos, et J.-M. Montel. 2002. « U-Th-Pb Dating of Phosphate Minerals ». *Reviews in Mineralogy and Geochemistry* 48 (1): 524-58. https://doi.org/10.2138/rmg.2002.48.14.

Haydock, Roger, et David R Kingham. s. d. *FIELD IONIZATION THEORY: A NEW, ANALYTIC, FORMALISM*.

Houard, J., A. Normand, E. Di Russo, et al. 2020. « A Photonic Atom Probe Coupling 3D Atomic Scale Analysis with *in Situ* Photoluminescence Spectroscopy ». *Review of Scientific Instruments* 91 (8): 083704. https://doi.org/10.1063/5.0012359.

Hudson, D., G.D.W. Smith, et B. Gault. 2011. « Optimisation of Mass Ranging for Atom Probe Microanalysis and Application to the Corrosion Processes in Zr Alloys ». *Ultramicroscopy* 111 (6): 480-86. https://doi.org/10.1016/j.ultramic.2010.11.007.

Jakob, Severin, Andrea Fazi, et Mattias Thuvander. 2024. « Laser-Assisted Field Evaporation of Chromia with Deep Ultraviolet Laser Light ». *Microscopy and Microanalysis*, novembre 14, ozae111. https://doi.org/10.1093/mam/ozae111.

Karahka, M., Y. Xia, et H. J. Kreuzer. 2015. « The Mystery of Missing Species in Atom Probe Tomography of Composite Materials ». *Applied Physics Letters* 107 (6): 062105. https://doi.org/10.1063/1.4928625.

Karahka, Markus, et H.J. Kreuzer. 2015. « Field Evaporation of Insulators and Semiconductors: Theoretical Insights for ZnO ». *Ultramicroscopy* 159 (décembre): 156-61. https://doi.org/10.1016/j.ultramic.2015.03.011.

Kingham, David R. 1982. « The post ionisation of field evaporated ions : A therorical explanation of multiple charge states ». *Surface Science 116*, janvier 21.

Kirubanithy, M., A. Albert Irudayaraj, A. Dhayal Raj, et S. Manikandan. 2015. « Synthesis, Characterization and Photoluminescence Behaviours of CePO4 and Tb-Doped CePO4 Nanostructures ». *Materials Today: Proceedings* 2 (9): 4344-47. https://doi.org/10.1016/j.matpr.2015.10.024.

Klaes, Benjamin, Rodrigue Lardé, Fabien Delaroche, et al. 2021. « Development of Wide Field of View Three-Dimensional Field Ion Microscopy and High-Fidelity Reconstruction Algorithms to the Study of Defects in Nuclear Materials ». *Microscopy and Microanalysis* 27 (2): 365-84. https://doi.org/10.1017/S1431927621000131.

Kumar, A, J Bogdanowicz, J Demeulemeester, et al. 2024. « Measurement of the Apex Temperature of a Nanoscale Semiconducting FIeld Emitter Illuminated by a Femtosecond Pulsed Laser ». *J. Appl. Phys.*

Kwak, Chang-Min, Young-Tae Kim, Chan-Gyung Park, et Jae-Bok Seol. 2017. « Understanding of Capping Effects on the Tip Shape Evolution and on the Atom Probe Data of Bulk LaAlO$_3$ Using Transmission Electron Microscopy ». *Microscopy and Microanalysis* 23 (2): 329-35. https://doi.org/10.1017/S1431927617000149.

Larson, D.J., T.J. Prosa, J.H. Bunton, et al. 2013. « Improved Mass Resolving Power and Yield in Atom Probe Tomography ». *Microscopy and Microanalysis* 19 (S2): 994-95. https://doi.org/10.1017/S143192761300696X.

Lefebvre-Ulrikson, Williams, éd. 2016. *Atom Probe Tomography: Put Theory into Practice*. Academic Press.

Morris et al. 2024. *Significant Oxygen Underestimation When Quantifying Barium-Doped SrTiO Layers by Atom Probe Tomography*.

Morris, Richard. J. H., Ramya Cuduvally, Davit Melkonyan, et al. 2018. « Toward Accurate Composition Analysis of GaN and AlGaN Using Atom Probe Tomography ». *Journal of Vacuum Science & Technology B, Nanotechnology and Microelectronics: Materials, Processing, Measurement, and Phenomena* 36 (3): 03F130. https://doi.org/10.1116/1.5019693.

Müller, Erwin W. 1951. « Das Feldionenmikroskop ». *Zeitschrift für Physik* 131 (1): 136-42. https://doi.org/10.1007/BF01329651.

Müller, Erwin W, John A Panitz, et Mclane S. Brooks. 1968. « The Atom-Probe Field Ion Microscope ». *The Review of Scientific Instruments* 39 (1): 83-85. https://doi.org/10.1063/1.478517.

Ndiaye, Samba, Christian Bacchi, Benjamin Klaes, Mariaconcetta Canino, François Vurpillot, et Lorenzo Rigutti. 2023. « Surface Dynamics of Field Evaporation in Silicon Carbide ». *The Journal of Physical Chemistry C* 127 (11): 5467-78. https://doi.org/10.1021/acs.jpcc.2c08908.

Ni, Yunxiang, John M. Hughes, et Anthony N. Mariano. 1995. « Crystal chemistry of the monazite and xenotime structures ». *American Mineralogist* 80 (1-2): 21-26. https://doi.org/10.2138/am-1995-1-203.

Parrish, Randall R. 1990. « U–Pb Dating of Monazite and Its Application to Geological Problems ». *Canadian Journal of Earth Sciences* 27 (11): 1431-50. https://doi.org/10.1139/e90-152.

Prosa, Ty J., et David J. Larson. 2017. « Modern Focused-Ion-Beam-Based Site-Specific Specimen Preparation for Atom Probe Tomography ». *Microscopy and Microanalysis* 23 (2): 194-209. https://doi.org/10.1017/S1431927616012642.

Reddy, Steven M., David W. Saxey, William D. A. Rickard, et al. 2020. « Atom Probe Tomography: Development and Application to the Geosciences ». *Geostandards and Geoanalytical Research* 44 (1): 5-50. https://doi.org/10.1111/ggr.12313.

Santhanagopalan, Dhamodaran, Daniel K. Schreiber, Daniel E. Perea, et al. 2015. « Effects of Laser Energy and Wavelength on the Analysis of LiFePO4 Using Laser Assisted Atom Probe Tomography ». *Ultramicroscopy* 148 (janvier): 57-66. https://doi.org/10.1016/j.ultramic.2014.09.004.

Schiester, Maximilian, Helene Waldl, Marcus Hans, et al. 2024. « Influence of Multiple Detection Events on Compositional Accuracy of TiN Coatings in Atom Probe Tomography ». *Surface and Coatings Technology* 477 (février): 130318. https://doi.org/10.1016/j.surfcoat.2023.130318.

Schwarz, Tim M, Eric Woods, Mahander P Singh, et al. 2024. « In Situ Metallic Coating of Atom Probe Specimen for Enhanced Yield, Performance, and Increased Field-of-View ». *Microscopy and Microanalysis* 00: 1-15. https://doi.org/10.1093/mam/ozae006.

Schwarz, Tim M, Eric Woods, Mahander P Singh, et al. s. d. *In-Situ Metallic Coating of Atom Probe Specimen for Enhanced Yield, Performance, and Increased Field-of-View*.

Sen, Amrita, Mukesh Bachhav, Francois Vurpillot, et al. 2021. « Influence of Field Conditions on Quantitative Analysis of Single Crystal Thorium Dioxide by Atom Probe Tomography ». *Ultramicroscopy* 220 (janvier): 113167. https://doi.org/10.1016/j.ultramic.2020.113167.

Sévelin-Radiguet, N., L. Arnoldi, F. Vurpillot, A. Normand, B. Deconihout, et A. Vella. 2015. « Ion Energy Spread in Laser-Assisted Atom Probe Tomography ». *EPL (Europhysics Letters)* 109 (3): 37009. https://doi.org/10.1209/0295-5075/109/37009.

Seydoux-Guillaume, A.-M., D. Fougerouse, A.T. Laurent, E. Gardés, S.M. Reddy, et D.W. Saxey. 2019. « Nanoscale Resetting of the Th/Pb System in an Isotopically-Closed Monazite Grain: A Combined Atom Probe and Transmission Electron Microscopy Study ». *Geoscience Frontiers* 10 (1): 65-76. https://doi.org/10.1016/j.gsf.2018.09.004.

Seydoux-Guillaume, Anne-Magali, Bernard Bingen, Valérie Bosse, Emilie Janots, et Antonin T. Laurent. 2018. « Transmission Electron Microscope Imaging Sharpens Geochronological Interpretation of Zircon and Monazite ». In *Geophysical Monograph Series*, 1re éd., édité par Desmond E. Moser, Fernando Corfu, James R. Darling, Steven M. Reddy, et Kimberly Tait. Wiley. https://doi.org/10.1002/9781119227250.ch12.

Silaeva, E. P., L. Arnoldi, M. L. Karahka, et al. 2014. « Do Dielectric Nanostructures Turn Metallic in High-Electric Dc Fields? » *Nano Letters* 14 (11): 6066-72. https://doi.org/10.1021/nl502715s.

Takahashi, Jun, Kazuto Kawakami, Koyo Miura, Mitsuhiro Hirano, et Naofumi Ohtsu. 2022. « Quantitative Analysis of Nitrogen by Atom Probe Tomography Using Stoichiometric $\Gamma'$-Fe$_4$N Consisting of $^{15}$N Isotope ». *Microscopy and Microanalysis* 28 (1): 42-52. https://doi.org/10.1017/S1431927621013623.

Tegg, Levi, Leigh T Stephenson, et Julie M Cairney. 2024. « Estimation of the Electric Field in Atom Probe Tomography Experiments Using Charge State Ratios ». *Microscopy and Microanalysis* 30 (3): 466-75. https://doi.org/10.1093/mam/ozae047.

Torkornoo, Selase, Marc Bohner, Ingrid McCarroll, et Baptiste Gault. 2024. « Optimization of Parameters for Atom Probe Tomography Analysis of β-Tricalcium Phosphates ». *Microscopy and Microanalysis*, août 30, ozae077. https://doi.org/10.1093/mam/ozae077.

Turuani, M J, A-M Seydoux-Guillaume, S M Reddy, et al. s. d. *Nanoscale Features Revealed by a Multiscale Characterization of Discordant Monazite Highlight Mobility Mechanisms of Th and Pb*.

Turuani, Marion. s. d. « Résoudre les perturbations des systèmes isotopiques U-Th-Pb à l'échelle nanométrique dans les monazites de contextes de UHT (Antarctique, Madagascar, Inde) ».

Turuani, M.J., A.T. Laurent, A.-M. Seydoux-Guillaume, et al. 2022. « Partial Retention of Radiogenic Pb in Galena Nanocrystals Explains Discordance in Monazite from Napier Complex (Antarctica) ». *Earth and Planetary Science Letters* 588 (juin): 117567. https://doi.org/10.1016/j.epsl.2022.117567.

Turuani, M.J., A.-M. Seydoux-Guillaume, A.T. Laurent, et al. 2024. « From ID-TIMS U-Pb Dating of Single Monazite Grain to APT-Nanogeochronology: Application to the UHT Granulites of Andriamena (North-Central Madagascar) ». *BSGF - Earth Sciences Bulletin* 195: 18. https://doi.org/10.1051/bsgf/2024013.

Valderrama, Billy, Hunter B. Henderson, Clarissa A. Yablinsky, Jian Gan, Todd R. Allen, et Michele V. Manuel. 2015. « Investigation of Material Property Influenced Stoichiometric Deviations as Evidenced during UV Laser-Assisted Atom Probe Tomography in Fluorite Oxides ». *Nuclear Instruments and Methods in Physics Research Section B: Beam Interactions with Materials and Atoms* 359 (septembre): 107-14. https://doi.org/10.1016/j.nimb.2015.07.048.

Vella, A., et J. Houard. 2016. « Chapter Eight - Laser-Assisted Field Evaporation ». In *Atom Probe Tomography*, édité par Williams Lefebvre-Ulrikson, François Vurpillot, et Xavier Sauvage. Academic Press. https://doi.org/10.1016/B978-0-12-804647-0.00008-5.

Verberne, Rick, Steven M. Reddy, Denis Fougerouse, et al. 2024. « Clustering and Interfacial Segregation of Radiogenic Pb in a Mineral Host-Inclusion System: Tracing Two-Stage Pb and Trace Element Mobility in Monazite Inclusions in Rutile ». *American Mineralogist* 109 (9): 1578-90. https://doi.org/10.2138/am-2023-9085.

Vurpillot, Francois, Stefan Parviainen, Fluyra Djurabekova, David Zanuttini, et Benoit Gervais. 2018. « Simulation Tools for Atom Probe Tomography: A Path for Diagnosis and Treatment of Image Degradation ». *Materials Characterization* 146 (décembre): 336-46. https://doi.org/10.1016/j.matchar.2018.04.024.

Yang, Wei, Yang-Ting Lin, Jian-Chao Zhang, Jia-Long Hao, Wen-Jie Shen, et Sen Hu. 2012. « Precise Micrometre-Sized Pb-Pb and U-Pb Dating with NanoSIMS ». *Journal of Analytical Atomic Spectrometry* 27 (3): 479. https://doi.org/10.1039/c2ja10303f.

Zanuttini, David, Ivan Blum, Lorenzo Rigutti, et al. 2017. « Simulation of Field-Induced Molecular Dissociation in Atom-Probe Tomography: Identification of a Neutral Emission Channel ». *Physical Review A* 95 (6): 061401. https://doi.org/10.1103/PhysRevA.95.061401.